\documentclass[]{interact}

\usepackage{epstopdf}
\usepackage[caption=false]{subfig}

\usepackage[numbers,sort&compress]{natbib}
\bibpunct[, ]{[}{]}{,}{n}{,}{,}
\makeatletter
\def\NAT@def@citea{\def\@citea{\NAT@separator}}
\makeatother

\theoremstyle{plain}

\theoremstyle{definition}

\theoremstyle{remark}

\begin{document}

\articletype{REVIEW}

\title{Optical Spatial Shock Waves in Nonlocal Nonlinear Media}

\author{
\name{Giulia~Marcucci\textsuperscript{a,b}\thanks{CONTACT Giulia~Marcucci. Email: giulia.marcucci@uniroma1.it}, Davide~Pierangeli\textsuperscript{b,a}, Silvia~Gentilini\textsuperscript{b,a}, Neda~Ghofraniha\textsuperscript{b,a}, Zhigang~Chen\textsuperscript{c,d} and Claudio~Conti\textsuperscript{b,a,c}}
\affil{\textsuperscript{a}Department of Physics, Sapienza University, Piazzale Aldo Moro 5, 00185 Rome, Italy;
\textsuperscript{b}Institute for Complex Systems, National Research Council, Via dei Taurini 19, 00185 Rome, Italy;
\textsuperscript{c}The Key Laboratory of Weak-Light Nonlinear Photonics, TEDA Applied Physics Institute, Nankai University, Tianjin 300457, China;
\textsuperscript{d}Department of Physics and Astronomy, San Francisco State University, San Francisco, California 94132, USA.}
}

\maketitle

\begin{abstract}

Dispersive shock waves are fascinating phenomena occurring when nonlinearity overwhelms linear effects, such as dispersion and diffraction.  Many features of shock waves are still under investigation, as the interplay with noninstantaneity in temporal pulses transmission and nonlocality in spatial beams propagation.

Despite the rich and vast literature on nonlinear waves in optical Kerr media, spatial dispersive shock waves in nonlocal materials deserve further attention for their unconventional properties.  Indeed, they have been investigated in colloidal matter, chemical physics and biophotonics, for sensing and control of extreme phenomena.

Here we review the last developed theoretical models and recent optical experiments on spatial dispersive shock waves in nonlocal media. Moreover, we discuss observations in novel versatile materials relevant for soft matter and biology.

\end{abstract}






\section{Introduction}\label{intro}

Dispersive shock waves~(DSWs) are rapidly oscillating solutions of hyperbolic partial differential equations that contrast the generation of multivalued regions through the formation of undular bores~\cite{1974Gurevich,1999Whitham,2005El,2007Barsi,2007HoeferPhysD,2007HoeferPRL,2012Crosta,2014Conforti,2014Moro,2016El,2016ElSmyth,2017Marcucci}. This class of phenomena was investigated in several physical fields, initially in shallow water waves~\cite{1966Peregrine} and ion-acoustic waves~\cite{1970Taylor}, then in oceanography~\cite{1988Smyth}, pulses propagation in photonic fibers~\cite{1989Rothenberg,2016Wetzel}, Bose-Einstein condensates~\cite{2004Damski,2004Kamchatnov,2004Perez,2005Simula,2006Hoefer,2008Chang}, quantum liquids~\cite{2006Bettelheim}, photorefractive media~\cite{2007El}, plasma physics~\cite{2008Romagnani}, viscous fluids~\cite{2016Maiden}, and diffracting optical beams~\cite{1967Akhmanov,2007HoeferPhysD,2007Ghofraniha,2007Wan,2008Conti,2012Ghofraniha,2013Garnier,2013Gentilini,2014Gentilini,2014Smith1,2015GentiliniPRA,2015GentiliniSciRep,2015Xu,2016Braidotti,2016Xu,2017Zannotti,2019Gautam}.

In 1967 Gardner, Greene, Kruskal and Miura developed a method to solve the Korteweg-de Vries~(KdV) equation, called inverse scattering transform~(IST)~\cite{1967Gardner}. Among all the equations solvable by IST, which allowed to find the mathematical formulation of exact solutions of such nonlinear models, KdV and the nonlinear Schr\"odinger equation~(NLSE) belong to the case with dispersive regularization of the aforementioned multivalued singularity. NLSE is a universal model that describes many phenomena, in particular a third-order nonlinear phenomenon in optics: the Kerr effect~\cite{2015Fibich}, a refractive index perturbation linearly scaling with the light intensity. Kerr effect can be generalized to the nonlocal case when the nonlinear response in a specific point depends on entire beam transverse profile. This occurs, e.g., in thermal media~\cite{1984Carter,2002Bang,2002Peccianti,2004Conti,2004Guo,2005Rotschild,2005Yakimenko,2007Barsi,2007Ghofraniha,2007Minovich,2010Folli,2012Folli,2012Maucher,2013Folli,2013Gentilini,2014Smith,2015Alberucci,2016Alberucci,2016ElSmyth}. In these materials, light propagation is affected by a highly nonlocal Kerr nonlinearity, ruled by nonlocal NLSE.

Unfortunately, IST is still of little use for the nonlocal NLSE and other theoretical approaches must be conceived, despite some recent progress in 2D media~\cite{2017Hokiris,2019Hokiris}. For many years, Whitham modulation and hydrodynamic approximation have predominated in solving the nonlocal NLSE~\cite{1999Whitham,2007Ghofraniha}. However, hydrodynamic approximation cannot describe light propagating beyond the shock point, and two new techniques have been developed, one coming from nuclear physics, the time asymmetric quantum mechanics~(TAQM)~\cite{1981Bohm,1989Bohm,1996Bollini,1997Castagnino,1998Bohm,1999Bohm,2002Delamadrid,2003Chruscinski,2004Chruscinski,2004Civitarese,2016Marcucci,2017Marcucci}, which models the nonlinear wave intrinsically irreversible propagation as a superposition of decaying resonances, and the wave turbulence theory~\cite{2007Picozzi,2012Can, 2013Garnier, 2014Picozzi, 2015Xu, 2016Xu, 2017Fusaro}, which uses a statistical interpretation of nonlinear optics.

This review aims to summarize all the current theoretical models to describe wave breaking of nonlocal NLSE solutions in diffracting optical beam propagation, and to highlight some of the most recent experimental observations of DSWs in spatial nonlinear photonics.

After an introductory section about the derivation of nonlocal NLSE in Sec.~\ref{nonlocalnlse}, we report the main theoretical approaches and results related to DSWs. Section~\ref{kerr} explains in details the difference between the wave breaking due to local Kerr effect, which causes shock both in phase and in intensity, and the one in nonlocal Kerr media, where the beam intensity follows the phase singularity adiabatically~\cite{2007Ghofraniha}. The most recent theoretical models of nonlinear wave propagation in highly nonlocal nonlinear media are treated in Secs.~\ref{TAQM}, \ref{turbulence}. Section~\ref{TAQM} treats DSWs generated by laser beams and gives an analytical description of their intrinsic irreversibility, due to the complexity of the dynamics rather than losses~\cite{2017Marcucci}. Section~\ref{turbulence} illustrates the importance of nonlocality in random dispersive waves nonlinear interaction to produce giant collective incoherent shock waves~\cite{2014Picozzi, 2015Xu, 2016Xu}.

The second part of the manuscript is a collection of experiments on DSW generation in thermal media. Sec.~\ref{exp} reports observations in Rhodamine solutions~\cite{2007Ghofraniha}. Output beam intensity profiles in Sec.~\ref{rhodamine} are modeled by TAQM both in two dimensional experiments, where decaying states describe the longitudinal propagation~\cite{2015GentiliniPRA,2015GentiliniSciRep}, and in the one dimensional approximation, having the proof that TAQM is an excellent approach also to analyze transverse intensity profiles beyond the shock point~\cite{2016Braidotti}.
The interplay of nonlinearity and disorder is illustrated in Sec.~\ref{disorder}. There, observations in Rhodamine with silica spheres~\cite{2012Ghofraniha} and in silica aerogel~\cite{2014Gentilini} exhibit the competition between randomness and nonlocal Kerr effect.
DSW generation processes in chemical~\cite{2014Smith1} and biological solutions~\cite{2019Gautam} are illustrated in Sec.~\ref{bio}.
Last but not least, Sec.~\ref{incoherent} shows recent experiments on the transition from coherent shocklets to a giant incoherent DSW in a photon fluid, modeled by wave turbulence theory~\cite{2015Xu}. 

%
%
\section{The Nonlocal Nonlinear Schr\"odinger Equation}\label{nonlocalnlse}

From Maxwell's equations, considering a region with zero charge, current and magnetization, we obtain the following electric field wave equation
\begin{equation}
-\nabla^2\mathbf{E} + \frac{1}{c^2}\partial_t^2\mathbf{E}=-\frac{1}{\epsilon_0c^2}\partial_t^2\mathbf{P},
\label{eq:wave}
\end{equation}
with $\mathbf{E}$ the electric field and $\mathbf{P}$ the medium nonlinear polarization~\cite{2008Boyd}.

The relation between $\mathbf{P}$ and $\mathbf{E}$ depends on the material properties. Including all the nonlinear terms, we have
\begin{equation}
\mathbf{P}=\epsilon_0\Bigr(\chi^{(1)}\mathbf{E}+\chi^{(2)}\mathbf{E}\mathbf{E}+\chi^{(3)}\mathbf{E}\mathbf{E}\mathbf{E}+\dots\Bigl)=\mathbf{P}^{(L)} + \mathbf{P}^{(NL)},
\label{eq:pol}
\end{equation}
where $1+\chi^{(1)}=n_0^2$, $n_0$ is the medium refractive index, $\chi^{(2)}$ and $\chi^{(3)}$ are tensors denoted as second and third order susceptibility, respectively.

One must take into account the temporal delay between the instant when the electric field reaches the medium and the medium response. For this reason, this radiation-matter interaction is more properly represented by the following non instantaneous superposition of linear and nonlinear polarization~\cite{2008Boyd}:
\begin{equation}
\mathbf{P}=\epsilon_0\Bigr(\chi^{(1)}\ast\mathbf{E}(t)+\chi^{(2)}\ast\mathbf{E}\mathbf{E}(t)+\chi^{(3)}\ast\mathbf{E}\mathbf{E}\mathbf{E}(t)+\dots\Bigl)=\mathbf{P}^{(L)} + \mathbf{P}^{(NL)},
\label{eq:pol2}
\end{equation}
where $\ast$ is the convolution product
$$\chi^{(n)}\ast\mathbf{E}\dots\mathbf{E}(t)=\int_{-\infty}^t \mathrm{d}t_1\int_{-\infty}^{t}\mathrm{d}t_2\dots\int_{-\infty}^{t}\mathrm{d}t_n\chi^{(n)}(t-t_1,\dots)\mathbf{E}(\mathbf{R},t_1)\dots\mathbf{E}(\mathbf{R},t_n).$$
If we have a third-order isotropic and centrosymmetric material, the nonlinear polarization is
\begin{equation}
\mathbf{P}^{(NL)}(\mathbf{R},t)=\epsilon_0\int^t_{-\infty}\mathrm{d}t_1\int^t_{-\infty}\mathrm{d}t_2\int^t_{-\infty}\mathrm{d}t_3\chi^{(3)}(t-t_1,t-t_2,t-t_3)\mathbf{E}(\mathbf{R},t_1)\mathbf{E}(\mathbf{R},t_2)\mathbf{E}(\mathbf{R},t_3)
\label{eq:pol3}
\end{equation}
and the related dielectric tensor changes as 
\begin{equation}
\epsilon_{new}=\epsilon + \epsilon_2\langle\mathbf{E}\cdot \mathbf{E}\rangle,
\label{epsilont}
\end{equation}
where $\langle\mathbf{E}\cdot \mathbf{E}\rangle=\frac{1}{2}|\mathbf{E}|^2$ is the square of the electric field time average.
The final refractive index causes the Kerr effect~\cite{2008Boyd}, a phenomenon that consists in a perturbation of the medium refractive index, proportional to the field intensity:
\begin{equation}
n=\sqrt{\epsilon_{new}}=\sqrt{\epsilon + \epsilon_2\langle\mathbf{E}\cdot \mathbf{E}\rangle}\approx n_0+n_2I,
\label{nkerr}
\end{equation}
with $I=|\mathbf{E}|^2$ the field intensity and $n_2$ the Kerr coefficient. 

The nonlocal Kerr effect is a third-order phenomenon, but the radiation-matter interaction depends on the whole intensity profile, as occurs in thermal media. In these materials, when an optical beam propagates, it locally heats the medium, and the resulting temperature gradient generates a variation of the density distribution and a refractive index perturbation~\cite{2007Ghofraniha,2007Minovich,2015Alberucci}:
\begin{equation}
\Delta n =\left(\frac{\partial n}{\partial T}\right)_0 \Delta T,
\end{equation}
with $\left(\frac{\partial n}{\partial T}\right)_0$ the thermo-optic coefficient of the sample at the steady-state.
It turns out that the nonlinear response induced at a specific spatial point is carried away to the surrounding region, and the size of this extended region determines the range of nonlocality. The heat conduction in optical thermal materials was termed ``response with an infinite range of nonlocality"~\cite{2005Rotschild} until 2007, when A. Minovich \textit{et al.}~\cite{2007Minovich} proved experimentally and theoretically that the nonlocal response of thermal optical media can be accurately described by a localized well function dependent only on the sample geometry, and not on the nature of the material.
This property allows to express the temperature variation, in a stationary limit, as governed by the following 3D heat equation~\cite{1984Carter,1989Wetterer,1997Brochard,2005Yakimenko,2005Rotschild,2007Ghofraniha,2007Minovich,2015Alberucci} with constant boundary conditions (at room temperature):
\begin{equation}
\left(\partial^2_X+\partial^2_Y+\partial^2_Z\right)\Delta T(\mathbf{R})=-\gamma|\mathbf{E}(\mathbf{R})|^2,
\label{eq:temperature0}
\end{equation}
where $\gamma=(L_{loss}\rho_0c_P D_T)^{-1}$, $L_{loss}$ is the loss characteristic length, $\rho_0$ is the material density, $c_P$ is the specific heat at constant pressure, $D_T$ is the thermal diffusivity and $\mathbf{R}=(X,Y,Z)=(\mathbf{R}_{\perp},Z)$. The solution can be written as
\begin{equation}
\Delta T(\mathbf{R}) =\int\int\int \mathrm{d}Z'\mathrm{d}\mathbf{R_{\perp}}'G(\mathbf{R_{\perp}}-\mathbf{R_{\perp}}',Z-Z')|\mathbf{E}(\mathbf{R_{\perp}}',Z')|^2,
\label{eq:temperature}
\end{equation}
with $G(\mathbf{R_{\perp}})$ a Green function that depends only on the sample geometry and the boundary conditions, and expresses the nonlocality of the nonlinear effect.
In principle, one can remove the $Z$-dependence by integrating along the longitudinal medium length $Z_0$~\cite{2007Minovich}, but we are interested in the Green function itself, and the longitudinal behavior of $G$ becomes as complicated as $Z_0$ becomes comparable to $L_{loss}$, getting smaller and strongly asymmetric near the boundaries~\cite{2015Alberucci}. Physically, the reason why this happens is due to the choice of heat equation to describe the nonlinear radiation-matter interaction: it works only in a neighborhood of the sample midpoint $\hat{Z}=Z_0/2$, not in proximity of the borders. Mathematically, this is deciphered in a longitudinal parabolic approximation with characteristic width $L_{nloc}=\sqrt{\frac{|n_2|}{\gamma\left|\frac{\partial n}{\partial T}\right|_0}}\propto\sqrt{L_{loss}}$:
\begin{equation}
\Delta T(\mathbf{R}) = \left[1-\frac{(Z-\hat{Z})^2}{2L_{nloc}}\right]\Delta T_{\perp}(\mathbf{R_{\perp}}).
\label{eq:parabolicT}
\end{equation}

From Eqs.~(\ref{eq:temperature0},\ref{eq:parabolicT}) we obtain the 2D heat equation
\begin{equation}
\left(\partial^2_X+\partial^2_Y\right)\Delta T_{\perp}(\mathbf{R_{\perp}})-L_{nloc}^{-2}\Delta T_{\perp}(\mathbf{R_{\perp}})=-\gamma I_{\perp}(\mathbf{R_{\perp}}),
\label{eq:temperature1}
\end{equation}
with $I_{\perp}(\mathbf{R_{\perp}})=\frac1{Z_0}\int \mathrm{d}Z|\mathbf{E}(\mathbf{R_{\perp}},Z)|^2$.
Eq.~(\ref{eq:temperature}) now reads
\begin{equation}
\Delta T_{\perp}(\mathbf{R_{\perp}})=\int\int \mathrm{d}\mathbf{R_{\perp}}'G_{\perp}(\mathbf{R_{\perp}}-\mathbf{R_{\perp}}')I_{\perp}(\mathbf{R_{\perp}}').
\label{eq:temperature2}
\end{equation}
In low absorption regime ($Z_0<<L_{loss}$) $\Delta T(\mathbf{R}) \sim \Delta T_{\perp}(\mathbf{R_{\perp}})$ and $\partial_Z I(\mathbf{R})\sim 0$ (intensity longitudinal changes are negligible as for solitary wave packets), therefore we attain $n[I](\mathbf{R})=n_0+\Delta n[I](\mathbf{R_{\perp}})$, with the refractive index nonlocal perturbation
\begin{equation}
\Delta n[I](\mathbf{R_{\perp}})=n_2 \int\int \mathrm{d}\mathbf{R_{\perp}}'K(\mathbf{R_{\perp}}-\mathbf{R_{\perp}}')I(\mathbf{R_{\perp}}'),
\label{eq:nonlinearity}
\end{equation}
and $n_2K(\mathbf{R_{\perp}})=\Bigl(\frac{\partial n}{\partial T}\Bigr)_0G_{\perp}(\mathbf{R_{\perp}})$.

By a comparison between the nonlocality length $L_{nloc}$ and the beam waist $W_0$, we can analyze two different limits: the standard Kerr effect in Eq.~(\ref{nkerr}) when $L_{nloc}\ll W_0$ (local approximation), i.e., $K(\mathbf{R_{\perp}}-\mathbf{R_{\perp}}')\sim\delta(\mathbf{R_{\perp}}-\mathbf{R_{\perp}}')$,
and the opposite case $L_{nloc}\gg W_0$, that is, the highly nonlocal approximation~(HNA), where $K\ast I(\mathbf{R})\sim K(\mathbf{R_{\perp}})P(Z)$, with $P(Z)=\int \mathrm{d}\mathbf{R_{\perp}}I(\mathbf{R})$ the power.

For a monocromatic field $\mathbf{E}(\mathbf{R},\tau)=\hat{\mathbf{E}}_0 A(\mathbf{R})e^{-\imath\omega \tau}$ in a third-order thermal medium, in paraxial and slowly varying envelope approximations, introducing the delayed time $\tau=t-\frac{n_0}{c}Z$ and adding a linear loss of characteristic length $L_{loss}$, from Eq.~(\ref{eq:wave}) we find that the propagation along $Z$ is ruled by the nonlocal NLSE~\cite{2007Ghofraniha}:
\begin{equation}
2\imath k \partial_Z A+\left(\partial^2_X+\partial^2_Y\right) A+2k^2 \frac{\Delta n[|A|^2]}{n_0} A=-\imath\frac{k}{ L_{loss}}A,
\label{eq:NLS1}
\end{equation}
with $k=\frac{2\pi n_0}{\lambda}=\frac{\omega n_0}c$ the wavenumber.

\subsection{Spatial Dispersive Shock Waves in Nonlocal Kerr Nonlinearity}\label{kerr}

Spatial DSWs are rapidly oscillating waves which regularize an abrupt discontinuity in phase through diffraction, that is, through the formation of intensity undular bores on the beam borders. Scientific community paid close attention to the theoretical description~\cite{1974Gurevich,1999Whitham,2006Hoefer} and experimental demonstration~\cite{1970Taylor,1989Rothenberg,2004Damski,2004Kamchatnov,2004Perez,2005Simula,2007Ghofraniha,2007Wan,2008Romagnani,2013Garnier,2016Maiden} of optical DSWs.
Here we summarize results on the defocusing DSWs in nonlocal media~\cite{2007Ghofraniha}. In such materials, the IST cannot describe the solutions, and we need other methods.

In next two sections, we detail two different methodologies for DSWs in nonlocal media: the TAQM~\cite{1981Bohm,1989Bohm,1996Bollini,1997Castagnino,1998Bohm,1999Bohm,2002Delamadrid,2003Chruscinski,2004Chruscinski,2004Civitarese,2016Marcucci,2017Marcucci} and the wave turbulence theory~\cite{2007Picozzi,2014Picozzi}. Both theories also prove that DSWs are intrinsically irreversible.

Starting from Eq.~(\ref{eq:NLS1}), through the scaling $x=\frac{X}{W_0}$, $y=\frac{Y}{W_0}$, $z=\frac{Z}{L}$, $\psi(x,y,z)=\frac{A(X,Y,Z)}{\sqrt{I_0}}$, with $I_0$ the intensity peak, $L=\sqrt{L_{nl}L_d}$, $L_{nl}=\frac{n_0}{k|n_2|I_0}$ the nonlinear length scale associated to a local Kerr effect, $L_d=kW_0^2$ the diffraction length, one obtains the normalized nonlocal NLSE
\begin{equation}
\imath \epsilon \partial_z \psi+\frac{\epsilon^2}{2}\left(\partial^2_x+\partial^2_y\right) \psi+\chi\theta\psi=-\imath\frac{\alpha}{2}\epsilon\psi,
\label{eq:NLS2}
\end{equation}
with $\epsilon=\frac{L_{nl}}{L}=\sqrt{\frac{L_{nl}}{L_d}}$ a small quantity in strongly nonlinear (or weakly diffracting) regime, as the one we are considering, $\chi=\frac{n_2}{|n_2|}$, $\theta=\left|\frac{k\Delta n L_{nl}}{n_0}\right|$, $\alpha=\frac L{L_{loss}}$.
From Eq.~(\ref{eq:temperature1})
\begin{equation}
-\sigma^2\left(\partial^2_x+\partial^2_y\right) \theta+\theta=|\psi|^2,
\label{eq:theta}
\end{equation}
where $\sigma=\frac{L_{nloc}}{W_0}$ is the nonlocality degree, which expresses the nature of the Kerr effect through the limits we have previously discussed: if $\sigma\ll1$ we are considering the local limit $\theta\sim|\psi|^2$, instead, if $\sigma\gg1$, by HNA $\theta\sim\kappa(x,y)p(z)$, with $p(z)=\int\mathrm{d}x\mathrm{d}y\left|\psi(x,y,z)\right|^2=\frac{P(Z)}{W_0^2I_0}$.

The fundamental laser mode (Gaussian TEM$_{00}$) is described by an axisymmetric Gaussian input $\psi_0(r)=\exp(-r^2)$, with $r=\sqrt{x^2+y^2}$, which evolves in the WKB approximation~\cite{1935Kemble} as $\psi(r,z)=\sqrt{\rho(r,z)}\exp\left[\imath\frac{\phi(r,z)}{\epsilon}\right]$. For $D=2$ the transverse dimensionality and $u=\partial_r\phi$ the phase chirp, from Eqs.~(\ref{eq:NLS2}, \ref{eq:theta}) one obtains
\begin{equation}
\begin{array}{rcl}
\partial_z\rho+\left[\frac{D-1}{r}\rho u+\partial_r(\rho u)\right]&=&-\alpha\rho,\\
\partial_zu+u\partial_ru-\chi\partial_r\theta&=&0,\\
-\sigma^2\left[\partial_r^2\theta+\frac{D-1}{r}\partial_r\theta)\right]+\theta&=&\rho.\\
\end{array}
\label{eq:WKB}
\end{equation}

\begin{figure}[h!]
\centering
\includegraphics[width=0.7\textwidth]{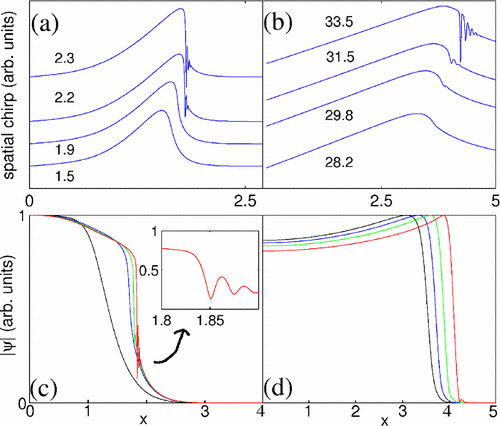}
\caption{Phase chirp $u(x)$ (a,b), and amplitude $\sqrt{\rho(x,z)}$ (c,d) for transverse dimensionality $D=1$ and different values of $z$, as indicated. (a,c) are obtained by Eqs.~(\ref{eq:WKBlocal}) with $\epsilon=10^{-3}$. (b,d) are simulations of the result of the system~(\ref{eq:WKB}) with $D=1,\;\alpha=0,\;\chi=-1,\;\sigma^2=5$. 

FIG.~1 reprinted with permission from~\cite{2007Ghofraniha}. Copyright 2007 by the American Physical Society.
\label{fig1}}
\end{figure}

Figure~\ref{fig1} reports phase chirp and field amplitude for $D=1$, so for $\partial_y\sim0$ and $r\rightarrow x$, in a defocusing medium $(\chi=-1)$ without losses $(\alpha=0)$.
The local case $(\sigma=0)$ is illustrated in Figs.~\ref{fig1}a,c and follows from system~(\ref{eq:WKB}):
\begin{equation}
\partial_z\rho+\partial_x(\rho u)=0,\;\;\;\partial_zu+u\partial_xu=-\partial_x\rho.
\label{eq:WKBlocal}
\end{equation}
Eqs.~(\ref{eq:WKBlocal}) are equivalent to Euler and continuity equations, respectively, for a fluid of speed $u$, mass density $\rho$ and pressure proportional to $\rho^2$.
In the reported dynamics, the diffraction, initially of order $\epsilon^2$, starts to play a relevant role in proximity of the wave breaking. In fact, it regularizes such a discontinuity by rapid oscillations of wavelength~$\sim\epsilon$, which appear simultaneously in phase chirp $u$ and intensity $\rho$.
For large values of $\sigma$, the normalized refractive index variation, here expressed by $\theta(x)$, is wider than the Gaussian input. As shown in Figs.~\ref{fig1}b,d the shock oscillations are essentially driven by the phase chirp $u$, while the intensity $\rho$ adiabatically follows.
Major details are given in~\cite{2007Ghofraniha}.

An in-depth description of the difference between DSWs in local and nonlocal Kerr media is also provided by turbulence theory, in particular by the Vlasov formalism, briefly summarized in Sec.~\ref{turbulence}. The analysis made for random optical waves in Sec.~\ref{turbulence} is also relevant to the coherent problem considered here, since the reduced hydrodynamic equations derived from the Vlasov model~[\cite{2015Xu}, Eqs. (3,4)] coincide with Eq.~(\ref{eq:WKB}). Following this approach, DSWs in thermal nonlinearity were interpreted for the first time as ``annular collapse singularities" in~\cite{2015Xu}. By looking at the M-shaped field amplitude in Fig.~\ref{fig1}d (and, in the following, at the intensity profiles in Fig.~\ref{fig5}a, Fig.~\ref{fig6}, Figs.~\ref{fig7}D,E, Figs.~\ref{fig9}a,b and Figs.~\ref{fig16}b,c), and comparing this to the fast oscillations in Fig.~\ref{fig1}c [or Fig.~2 in~\cite{2007Wan}], the feature of the collapse singularity in nonlocality appears evident. Indeed, the corresponding hydrodynamic model in the limit of a local nonlinearity [Eq.~(\ref{eq:WKBlocal})] recall the shallow water equations, which exhibit a pure shock without collapse.

\subsection{High Nonlocality and Time Asymmetric Quantum Mechanics}\label{TAQM}

Let us consider the nonlocal NLSE in Eq.~(\ref{eq:NLS1}) with a medium response function $K(X,Y)=\exp\left[-(|X|+|Y|)/L_{nloc}\right]/(2L_{nloc})^2$. Being $K$ separable, i.e., $K(X,Y)=\tilde K(X) \tilde K(Y)$, through the approximation $\partial_Y\sim0$ (as in the previous section) we can consider only one transverse dimension, since analyzing propagation along $Y$ is no more interesting for our purposes.
We rewrite Eq.~(\ref{eq:NLS1}) in terms of $1+1$ dimensionless variables by using the same scaling of Eq.~(\ref{eq:NLS2}) and choosing $I_0$ such that $L_{nl}=L_d$:
\begin{equation}
  \imath \partial_z \psi+\frac{1}{2}\partial_x^2 \psi-\kappa\ast|\psi|^2  \psi=-\imath\frac{\alpha}{2}\psi.
\label{paraxialnorm}
\end{equation}
with $\kappa(x)=W_0\tilde K(x W_0)=\exp\left(-|x|/\sigma\right)/(2\sigma)$.

We take into account a medium where the nonlocality length is much larger than the beam waist. By HNA we have \cite{1997Snyder,2012Folli}
\begin{equation}
\kappa*|\psi|^2\sim \kappa(x)p(z),
\label{eq:HNA}
\end{equation}
where $\kappa$ is a function no more depending on $|\psi|^2$.
In a system without loss, that is, $\alpha=0$, the normalized power $p$ is conserved and the NLSE is mapped into a linear Schr\"odinger equation $\imath \partial_z\psi=\hat{H}\psi$, with the Hamiltonian $\hat{H}=\tfrac{1}{2}\hat{p}^2+p \kappa(x)$ $\left(\hat{p}=-\imath\partial_x\right)$. When we express the even function $\kappa$ as its second order expansion, that is, $\kappa(x)=\kappa_0^2-\frac{\kappa_2^2}{2} x^2$, where $\kappa_0^2=\frac1{2 \sigma}$ and $\kappa_2^2=\frac1{\sqrt{\pi}\sigma^2}$, we obtain the reversed harmonic oscillator~(RHO) Hamiltonian~\cite{1999Bohm,2003Chruscinski,2016Marcucci}:
\begin{equation}
\label{eq:new_H}
\hat{H}=p\kappa_0^2 +\hat{H}_{RHO},\;\;\hat{H}_{RHO}=\frac{\hat{p}^2}{2}-\frac{\gamma^2 \hat{x}^2}{2},\;\;\gamma^2=p\kappa_2^2.
\end{equation}
If $\psi=\exp\left(-\imath \kappa_0^2 p z\right)\phi$, then $\imath\partial_z\phi=\hat{H}_{RHO}\phi$.

\begin{figure}[h!]
\centering
\includegraphics[width=0.9\textwidth]{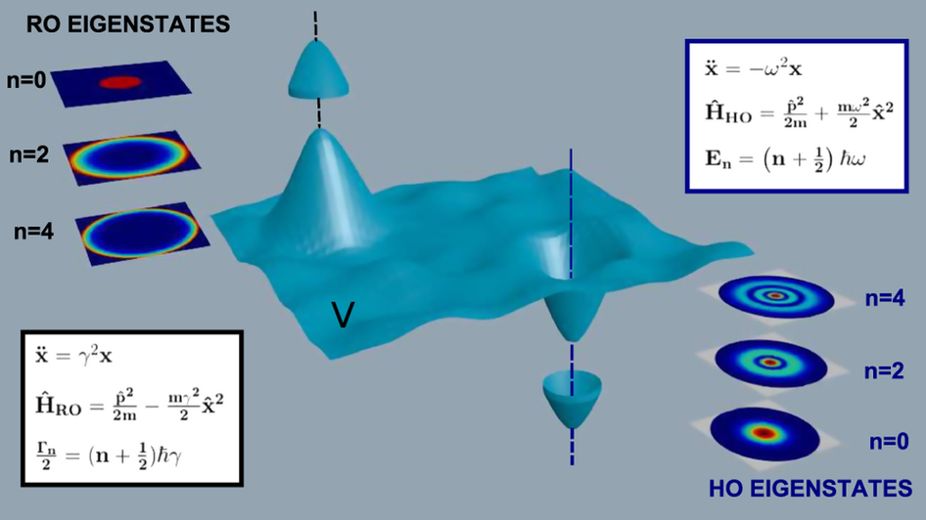}
\caption{
Pictorial representation of an energy landscape. When the system is in proximity of a local maximum it obeys the RHO Hamiltonian, in figure $\hat H_{RO}$. In proximity of the minimum the system obeys the Hamiltonian of a harmonic oscillator, in figure $\hat H_{HO}$. The two Hamiltonians are explicitly written in the two the corresponding text boxes, with the related dynamical systems and the discrete eingenvalues. Insets show the transverse profiles of the respective eigenfunctions, bounded on right hand side for the harmonic oscillator, unbounded on the left hand side for the RHO.

Reprinted by permission from Macmillan Publishers Ltd. from~\cite{2015GentiliniSciRep}. Copyright 2015.
\label{fig3}}
\end{figure}

Figure~\ref{fig3} sketches the relation between the harmonic and the reversed oscillators. For a harmonic oscillator~(HO), the spectrum is discrete and the corresponding eingenstates form a orthonormal basis (both a orthogonality and a completeness relations hold):
\begin{equation}
\begin{array}{l}
\label{HO}
\hat{H}_{HO}=\frac{\hat p^2}{2}+\frac{\omega^2}{2}\hat x^2\\
\hat{H}_{HO}\psi(x)=E\psi(x),\;\;E_n=\omega\left(n+\frac{1}{2}\right),\\
\psi_n(x)=\sqrt[4]{\frac{\omega}{\pi}}\frac{1}{\sqrt{2^nn!}}H_n\left(\sqrt{\omega}x\right),
\end{array}
\end{equation}
with $H_n(x)=(-1)^nx^2\frac{d^n}{dx^n}e^{-x^2}$ the Hermite polynomials.
On the other hand, RHO has complete continuous spectrum, but one derives a generalized discrete spectrum from HO spectrum by a complex analytic prolongation  in the rigged Hilbert space~\cite{1998Bohm,2016Celeghini,2016Marcucci} through the transformation $\omega\rightarrow\imath\gamma$, $\hat x\rightarrow e^{-\imath\frac\pi 4}\hat x$, $\hat p\rightarrow e^{\imath\frac\pi 4}\hat p$~\cite{2003Chruscinski,2016Marcucci}. The new stationary Schr\"odinger equation is $\hat{H}_{RHO}\mathfrak{f}^{\pm}(x)=\imath\frac{\Gamma}2\mathfrak{f}^{\pm}(x)$, solved by the spectrum $\frac{\Gamma_n}2=\gamma\left(n+\frac{1}{2}\right)$ and the non normalizable eigenfunctions
\begin{equation}
  \mathfrak{f}_n^\pm(x)=\frac{\sqrt[4]{\pm i \gamma}}{\sqrt{2^n n! \sqrt{\pi}}}  H_n(\sqrt{\pm i\gamma}x)\exp(\mp i\frac{\gamma}{2} x^2),
\label{eq:GVs}
\end{equation}
namely, the RHO \textit{Gamow vectors}~(GVs)~\cite{1928GamowDE,1928GamowENG}.

\begin{figure}[h!]
\centering
\includegraphics[width=0.75\textwidth]{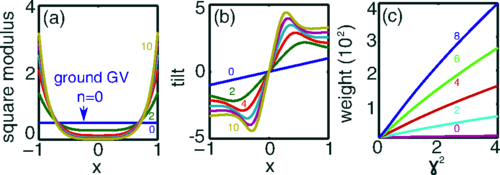}
\caption{
(a) $||\mathfrak{f}_n^-(x)||^2$ in Eq.~(\ref{eq:GVs}) for increasing even order $n$; (b) corresponding phase chirps $\partial_x\mathrm{Arg}\left[\mathfrak{f}_n^-(x)\right]$; (c) weights $p_n(0)$ [Eq.~(\ref{expprob})] of the GV expansion of a Gaussian wave packet.

FIG.~1 reprinted with permission from~\cite{2015GentiliniPRA}. Copyright 2015 by the American Physical Society.
\label{fig4}}
\end{figure}

We can express every wavefunction as a truncated superposition of GVs added to a background function, which dispersively oscillates at infinite as a polynomials~\cite{2016Marcucci}:
\begin{equation}
\phi(x)=\phi_N^G(x)+\phi_N^{BG}(x)
\label{eq:RHOGV}
\end{equation}
with
\begin{equation}
\phi_N^G(x)=\sum_{n=0}^N \mathfrak{f}_n^-(x) \langle\mathfrak{f}_n^+|\phi(x,0)\rangle.
\label{GamowComponents}
\end{equation}
Figure~\ref{fig4} shows the GV square norms (Fig.~\ref{fig4}a) and phase chirps (Fig.~\ref{fig4}b).
The evolution of the normalized field $\psi$ presents a Gamow part resulting as a superposition of exponential decays with quantized decay rates~\cite{2016Marcucci}: 
\begin{equation}
\psi_n^G(x,z)=\sum_{n=0}^N \langle \mathfrak{f}_n^+|\psi(x,0)\rangle \mathfrak{f}_n^-(x)e^{-i \kappa_0^2 p z} e^{-\frac{\Gamma_n}{2} z}.
\label{gamowevolution}
\end{equation}
Eq.~(\ref{gamowevolution}) proves an intrinsical irreversibility of DSWs, where a backward propagation beyond the shock point is no physically possible because of the exponentially decaying evolution. This explains why the quantum representation of wave propagation theory in a rigged Hilbert space is called TAQM (here time is replaced by $z$).

\begin{figure}[h!]
\centering
\includegraphics[width=0.75\textwidth]{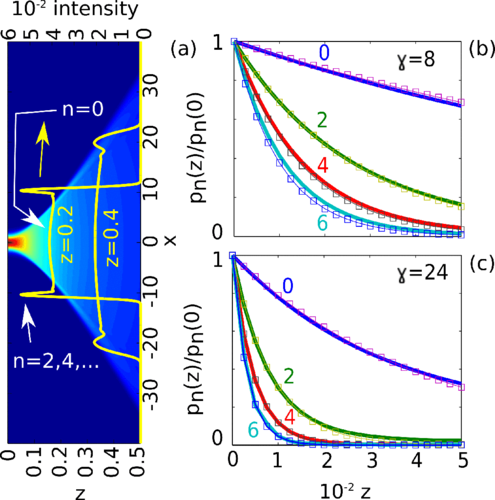}
\caption{
(a) Numerical solution of Eq.~(\ref{paraxialnorm}) with $p=10^4$ and $\sigma^2=10$; (b) projection on GVs for increasing order $n$ for $\alpha=0.3$ and $\gamma=8$; continuous lines are from Eq.~(\ref{paraxialnorm}), dots are from Eq.~(\ref{expprob}); (c) as in panel (b) for $\gamma=24$.

FIG.~2 reprinted with permission from~\cite{2015GentiliniPRA}. Copyright 2015 by the American Physical Society.
\label{fig5}}
\end{figure}

In the probabilistic interpretation of TAQM~\cite{2015GentiliniPRA}, the projection of Eq.~(\ref{gamowevolution}) over $\sqrt{\Gamma_n} \mathfrak{f}_n^+$ gives the probability $p_n(z)$ of finding the system in a decaying GV
\begin{equation}
p_n(z)= \Gamma_n |\langle \mathfrak{f}_n^+|\psi(x,0)\rangle|^2 e^{-\Gamma_n z},
\label{expprob}
\end{equation}
which gives the $z-$dependent weight of the $n$-order GV. Initial weights $p_n(0)$ are reported in Fig.~\ref{fig4}c as functions of $\gamma^2$.
Since a Gaussian beam  $\psi(x,0)=\varphi(x)=\exp(-x^2/2)/\sqrt[4]{\pi}$ is an even input, all the odd terms in Eq.~(\ref{gamowevolution}) vanish due the $x-$parity.
Figure~\ref{fig5}a shows the numerical solution of Eq.~(\ref{paraxialnorm}). Yellow lines give the transverse intensity profile. We see that these are modeled by a superposition of exponential decays, where the plateau is given by the groundstate GV, and the peaks are given by higher order GVs.
Simulations of weights $p_n(z)$ are in Figs.~\ref{fig5}b,c. While dotted profiles are numerical results from Eq.~(\ref{expprob}), continuous lines result from the general projection definition $p_n(z)= \Gamma_n |\langle \mathfrak{f}_n^+|\psi(x,z)\rangle|^2$, with $\psi(x,z)$ numerical solution of Eq.~(\ref{paraxialnorm}).

\subsection{Random Optical Waves and Turbulence Theory}\label{turbulence}

We have summarized how TAQM describes the intrinsic irreversibility of DSWs generated by a laser input. We highlight that nonlinear optics offers other examples of intrinsical irreversibility, as photon fluids, i.e., turbulent flows of a conservative system of random optical waves, in which the statistical interpretation has produced significant results~\cite{2014Picozzi}.

Statistical nonlinear optics is related to wave turbulence theory, whereby the kinetic wave description provides a thermodynamic treatment of turbulence~\cite{1992Zakharov,2001Newell,2007Picozzi,2011Nazarenko,2011Newell,2012Can,2013Garnier,2014Picozzi,2015Xu,2016Xu}.
If one considers the nonlinear propagation of incoherent optical waves characterized by fluctuations that are statistically inhomogeneous in space, DSWs arise only in the strong turbulence (strongly nonlinear) regime~\cite{2015Xu,2016Xu}.
The wave breaking emerges from a turbulent field, whose local averaged spectrum follows a specific Vlasov-like kinetic equation, which can be a traditional Vlasov equation (local nonlinearity), a short-range Vlasov equation~(SRVE) (quasi-local nonlinearity) or a long-range Vlasov equation~(LRVE) (highly nonlocal nonlinearity), derived by the zero-loss NLSE through a multiscale expansion~\cite{2014Picozzi}.
These Vlasov equations, being reversible kinetic equations, do not exhibit a H-theorem of entropy growth, and so do not describe the process of irreversible thermalization to equilibrium. They can be interpreted as a mean-field theory that does not explain, by itself, irreversible processes.
For the sake of completeness, we stress that the thermalization process arises at the next order of a weakly nonlinear expansion procedure. This analysis leads to a different kind of time asymmetric behavior with respect to DSWs reported above, because it is not due to strong nonlinearity. In this framework, by going to next order, the theory reveals that the Vlasov equation is corrected by a collision term involving the nonlocal interaction. This collision term has a form analogous to the wave turbulence kinetic equation~\cite{1992Zakharov} and is written in explicit form in Methods of~\cite{2015Xu}.
The mathematical statement of such irreversibility relies on the H-theorem of entropy growth. Wave thermalization can be characterized by a self-organization process, that is, the system spontaneously generates a large-scale coherent structure. A remarkable example of this self-organization process is the wave condensation, whose thermodynamic equilibrium properties are similar to those of quantum Bose-Einstein condensates.
We stress that nonlocality significantly decelerates the rate of the thermalization process, in complete analogy with gravitational systems, where the dynamics of stars are described by a collision-less (Vlasov-like) equation toward quasi-stationary nonequilibrium states (e.g. galaxies), which are of fundamental different nature than usual thermodynamic equilibrium states~\cite{2014Campa}.
A detailed treatise of these phenomena is reported in~\cite{2015Xu} and in the references therein.

In what follows, we analyze long-range interactions in strongly nonlinear wave systems operating far from thermodynamic equilibrium. 
Starting from the NLSE [Eq.~(\ref{eq:NLS1})], with nonlinearity expressed by Eq.~(\ref{eq:nonlinearity}) and no loss, we obtain Eq.~(\ref{eq:NLS2}), here explicitly written as
\begin{equation}
\imath \partial_z \psi+\frac{\epsilon}{2}\left(\partial^2_x+\partial^2_y\right) \psi+\chi\upsilon\psi U\ast|\psi|^2=0,
\label{eq:NLS3}
\end{equation}
 with $\upsilon=\frac{|n_2| I_0}{n_0}k\sqrt{L_{nl}L_d}$, $U(\mathbf{r})=W_0^2K(W_0\mathbf{r})$, $\mathbf{r}=(x,y)$.
From Eq.~(\ref{eq:NLS3}), we want to attain a kinetic equation, that is, an equation describing the evolution of the spectrum during the related field propagation in the nonlinear medium.
The structure of a kinetic equation depends on the statistics of the random wave. For this reason, we consider the field autocorrelation function
\begin{equation}
\begin{array}{ccc} B(\mathbf{r},\mathbf{\xi},z)=\langle\psi(\mathbf{r}+\frac12\mathbf{\xi},z)\psi^*(\mathbf{r}-\frac12\mathbf{\xi},z)\rangle, & \mathbf{r}=(\mathbf{r_1}+\mathbf{r_2})/2, & \mathbf{\xi}=\mathbf{r_1}-\mathbf{r_2}.\end{array} 
\label{eq:auto}
\end{equation} 
The statistics is said to be homogeneous if $B$ depends only on the distance $|\mathbf{r_1}-\mathbf{r_2}|$.
Following Eq.~(\ref{eq:NLS3})
\begin{equation}
\imath \partial_z B(\mathbf{r},\mathbf{\xi},z)+\epsilon\nabla_{\mathbf{r}}\cdot\nabla_{\mathbf{\xi}}B(\mathbf{r},\mathbf{\xi},z)+\chi\upsilon\left[P(\mathbf{r},\mathbf{\xi},z)-Q(\mathbf{r},\mathbf{\xi},z)\right]=0,
\label{eq:B}
\end{equation}
with
\begin{equation}
\begin{array}{rcl}
P(\mathbf{r},\mathbf{\xi},z)&=&B(\mathbf{r},\mathbf{\xi},z)\int \mathrm{d}\mathbf{r}' U(\mathbf{r}')\left[N(\mathbf{r}-\mathbf{r}'+\frac12\mathbf{\xi},z)-N(\mathbf{r}-\mathbf{r}'-\frac12\mathbf{\xi},z)\right],\\ \\
Q(\mathbf{r},\mathbf{\xi},z)&=&\int \mathrm{d}\mathbf{r}' U(\mathbf{r}')\left[B(\mathbf{r}-\frac12\mathbf{r}'+\frac12\mathbf{\xi},\mathbf{r}',z)B(\mathbf{r}-\frac12\mathbf{r}',\mathbf{\xi}-\mathbf{r}',z)+\right.\\
& &\left.-B(\mathbf{r}-\frac12\mathbf{r}',\mathbf{\xi}+\mathbf{r}',z)B(\mathbf{r}-\frac12\mathbf{r}'-\frac12\mathbf{\xi},-\mathbf{r}',z)\right],\\ \\
N(\mathbf{r},z)&=&B(\mathbf{r},\mathbf{\xi}=\mathbf{0},z)=\langle|\psi|^2\rangle(\mathbf{r},z).
\end{array}
\label{eq:PQN}
\end{equation}
In the last equation, $N$ is the averaged field power, also depending on $\mathbf{r}$ because of the inhomogeneity of the statistics. By defining the length scale of random fluctuations as the coherence length $\lambda_c$ and $W_0$ the incoherent beam waist, we can assume that the statistics is quasi-homogeneous if $\epsilon_c=\lambda_c/W_0<<1$.
For a Gaussian response function $U(\mathbf{r})=\frac1{2\pi\sigma^2}\exp\left(-\frac{|\mathbf{r}|^2}{2\sigma^2}\right)$, we get the SRVE if $\sigma<<1$ and the LRVE if $\sigma>>1$ through two different multiscale expansion with respect to $\epsilon_c$:
\begin{equation}
\partial_z n_{\mathbf{k}}(\mathbf{r},z)+\nabla_{\mathbf{k}}\tilde{\omega}_{\mathbf{k}}(\mathbf{r},z)\cdot\nabla_{\mathbf{r}}n_{\mathbf{k}}(\mathbf{r},z)-\nabla_{\mathbf{r}}\tilde{\omega}_{\mathbf{k}}(\mathbf{r},z)\cdot\nabla_{\mathbf{k}}n_{\mathbf{k}}(\mathbf{r},z)=0,
\label{eq:vlasov}
\end{equation}
for the local spectrum $n_{\mathbf{k}}(\mathbf{r},z)=\int\mathrm{d}\mathbf{\xi}B(\mathbf{r},\mathbf{\xi},z)\exp({-\imath\mathbf{k}\cdot\mathbf{\xi}})$,
with averaged power $N(\mathbf{r},z)=\frac1{(2\pi)^2}\int\mathrm{d}\mathbf{k}n_{\mathbf{k}}(\mathbf{r},z)$.
The nonlocal features of Eq.~(\ref{eq:vlasov}) are traced by the generalized dispersion relation. Once defined the Fourier trasform of the response function $\tilde{U}(\mathbf{k})=\int\mathrm{d}\mathbf{r}U(\mathbf{r},z)\exp({-\imath\mathbf{k}\cdot\mathbf{r}})$, for the LRTE
\begin{equation}
\begin{array}{rcl}
\tilde{\omega}_{\mathbf{k}}(\mathbf{r},z)&=&\omega(\mathbf{k})+V(\mathbf{r},z),\\ \\
\omega(\mathbf{k})&=&\frac{\epsilon}2|\mathbf{k}|^2,\\ \\
V(\mathbf{r},z)&=&-\chi\upsilon\int\mathrm{d}\mathbf{r}'U(\mathbf{r}-\mathbf{r}')N(\mathbf{r}',z).
\end{array}
\label{eq:dispersion}
\end{equation}
One can also compute the momentum both from the NLSE and the LRVE:
\begin{equation}
\begin{array}{rcl}
\mathbf{p}_{NLSE}(\mathbf{r},z)&=&(2\pi)^2\Im\left(\psi^*\nabla\psi\right)/N(\mathbf{r},z),\\ \\
\mathbf{p}_{LRVE}(\mathbf{r},z)&=&\int\mathrm{d}\mathbf{k}n_{\mathbf{k}}(\mathbf{r},z)\mathbf{k}/N(\mathbf{r},z).
\end{array}
\label{eq:momentum}
\end{equation}
Major details are given in~\cite{2014Picozzi}.

By increasing the range of the nonlocality, we pass from a stage where the field evolution is ruled by stochastic generation of small-scale DSW structures, naturally denoted as dispersive shocklets, to the emergence of an unexpected global collective behavior. The latter phenomenon is characterized by a strong non-homogeneous redistribution of the spatial fluctuations, whose description is provided by the NLSE and the LRVE, and unveils the formation of a giant shock singularity.
From theoretical analysis~\cite{2016Xu}, it turns out that - in the short-range regime - wave breaking occurs at random positions in the turbulent field, predominantly around high-amplitude fluctuations, leading to a gas of coherent dispersive shocklets in the midst of turbulent fluctuations.
On the other hand, for highly nonlocal Kerr samples, the regularization of the global incoherent shock does not require the formation of a regular oscillating DSW structure because the self-organization of the turbulent waves ensemble makes the field oscillate as a whole.
The momentum of the speckled beam [Eq.~(\ref{eq:dispersion})] is radially outgoing and exhibits a shock-like singularity, while the envelope of the intensity of the beam experiences an ring-shaped collapse-like behavior. The fluctuations of the incoherent wave then result to be pushed towards the annular shock front, which leaves behind itself an internal region of the beam with a high degree of coherence. In other terms, the dynamics is featured by a dramatic degradation of the coherence properties on the annular boundary of the beam, while its internal region exhibits a significant coherence enhancement.
Experimental observations of incoherent DSWs are reported in Sec.~\ref{incoherent}, together with related numerical simulations.

This alternation between coherence degradation on the boundaries and relative enhancement on the internal region also emerges in the spectrograms achievable by LRVE simulations. It turns out that, as the pump power increases, the spectrogram is affected by a Z-shaped distortion: the coherence length $\lambda_c\sim\Delta k^{-1}$ decreases at shock front ($\Delta k_{shock}$ determines the boundaries of the admitted $k$ values), and it increases in the internal region of the beam. This dynamics is conservative but irreversible, because the non-equilibrium thermodynamic condition is stabilized by an irreversible self-organization of the random waves. Indeed, in~\cite{2017Fusaro} authors showed that a spontaneous long-range phase coherence emerges among incoherent waves in nonlocal nonlinearity, when the speckles self-organize into giant collective waves. Their theory reveals that this phenomenon constitutes a generic property of a conservative (Hamiltonian) system of highly nonlocal random waves that evolve in a strongly nonlinear regime. Moreover, the field exhibits intensity fluctuations whose coherence length increases dramatically during the evolution of the system.

\section{Experimental Observations in Thermal Media}\label{exp}

\begin{figure}[h!]
\centering
\includegraphics[width=0.7\textwidth]{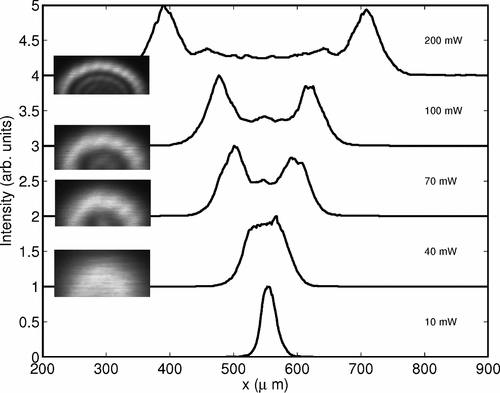}
\caption{
Experimental transverse intensity profiles of an initial Gaussian beam propagating in a thermal medium. Measurements are performed for varying input power $P=\pi W_0^2 I_0$. Insets show the 2D output patterns.

FIG.~5 reprinted with permission from~\cite{2007Ghofraniha}. Copyright 2007 by the American Physical Society.
\label{fig6}}
\end{figure}

Shock waves described by Eq.~(\ref{eq:NLS1}) have been originally shown in an experiment from~\cite{2007Ghofraniha}. The sample is a cell of length $1$mm  filled  with  an  aqueous solution of rhodamine B~(RhB), with a concentration of $0.6$mM. Measurements of the shock profiles are in Fig.~\ref{fig6}. A Gaussian CW laser beam of intensity waist $W_0=20\mu$m, at wavelength $\lambda=532$nm, propagates in a material with linear refractive index $n_0=1.3$,  defocusing  Kerr coefficient $n_2=-7\times10^{-7}$cm$^2$W$^{-1}$, loss length $L_{loss}^{-1}=62$cm$^{-1}$. For water $D_T=1.5\times10^{-7}$m$^2$s$^{-1}$, $\rho_0=10^3$kg m$^{-3}$, $c_p=4\times10^3$J kg$^{-1}$K$^{-1}$, $\left|\frac{\partial n}{\partial T}\right|_0=10^{-4}$K$^{-1}$. The degree of nonlocality is estimated as $\sigma=0.3$.

The beam exhibits, beyond the shock point, the formation of undular bores moving outward with increasing power. 
The next subsections report different experiments exhibiting DSWs in nonlocal samples.

\subsection{Rhodamine and Time Asymmetric Quantum Mechanics Interpretation}\label{rhodamine}

In this section we report two experiments in order to validate the presence of GVs in DSWs: a 2D propagation pattern to observe GVs decay rates $\Gamma_n$~\cite{2015GentiliniSciRep}, and a 1D experiment to show that GVs describe also the M-shaped profile in the far field of a DSW in HNA~\cite{2016Braidotti}, in~\cite{2015Xu} identified for the first time as collapse singularity. These are validations of TAQM in describing DSW propagation.

\begin{figure}[h!]
\centering
\includegraphics[width=0.9\textwidth]{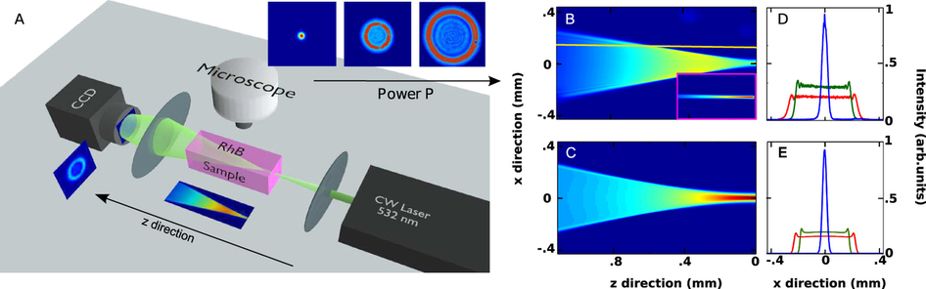}
\caption{
\textbf{A} Experimental setup. Authors of~\cite{2015GentiliniSciRep,2016Braidotti} collected the transmitted and fluorescence images of the laser beam propagating in RhB samples. Two types of launching lenses L1 were used: a cylindrical and a spherical, for the 1D and 2D experiments, respectively. The top fluorescence image of the propagating beam was collected by a microscope placed above the RhB samples. The second lens is spherical and was used to collect the transverse output profile.
\textbf{B,C} Top-view intensity distribution as obtained from 2D experiment B and numerical simulations C. Respectively experimental \textbf{D} and numerical \textbf{E} sections of the images B and C taken at $z = 0.2$ (red), $0.6$ (green) and $0.9$mm (blue).

Reprinted by permission from Macmillan Publishers Ltd. from~\cite{2015GentiliniSciRep}. Copyright 2015.
\label{fig7}}
\end{figure}

The experimental setup is illustrated in Fig.~\ref{fig7}A.
Samples are prepared by dispersing $0.1$mM of RhB in water. The solution is placed in a cuvette $1$mm thick in the pro\-pa\-ga\-tion direction. The measured defocusing Kerr coef\-fi\-cient is $|n_2|=2\times10^{-12}$m$^2$W$^{-1}$ and the absorption length is $L_{loss}\simeq1.6$mm at the laser wavelength $532$nm~\cite{2012Ghofraniha}.
The CW laser beam is focused through a lens into a sample. Light is collected by a spherical lens and a Charged Coupled Device~(CCD) camera. A microscope is placed above the sample in order to capture top-view images of the laser beam along the propagation direction $Z$. 
The difference between the two experimental apparatus is the choice of the first lens (L1).
In the 2D experiment~\cite{2015GentiliniSciRep}, L1 is spherical with focal length $10$cm, and a focus spot size of $10\mu$m. The setup was placed having the beam propagating vertically through the sample, reducing thermal convection in the water.
In the 1D expe\-ri\-ment~\cite{2016Braidotti}, authors used a cylindrical lens as L1, with focal length $f=20$cm in order to mimic a nearly one-dimensional propagation. Being $Z$ the propagation direction, the lens focuses the beam in the $X$ direction. The input spot dimension is $1.0$mm in the $Y$ direction and $35\mu$m in the $X$ direction. These geometrical features make the one-dimensional approximation valid and allow us to compare experimental results with the theoretical one-dimensional model. The diffraction length in the $X$ direction is $L_{d}=3.0$mm. This time, the setup was placed horizontally.

Figures~\ref{fig7}B,C report the observed laser beam pro\-pa\-ga\-tion top-view, detected by a microscope through RhB fluorescence, and the numerical calculation from the NLSE, respectively. The beam displays the characteristic strongly defocusing and the M-shaped behavior, also evident in the transverse sections of the intensity in Figs.~\ref{fig7}D,E. These are signatures of DSWs in nonlinear media at high power.

\begin{figure}[h!]
\centering
\includegraphics[width=0.8\textwidth]{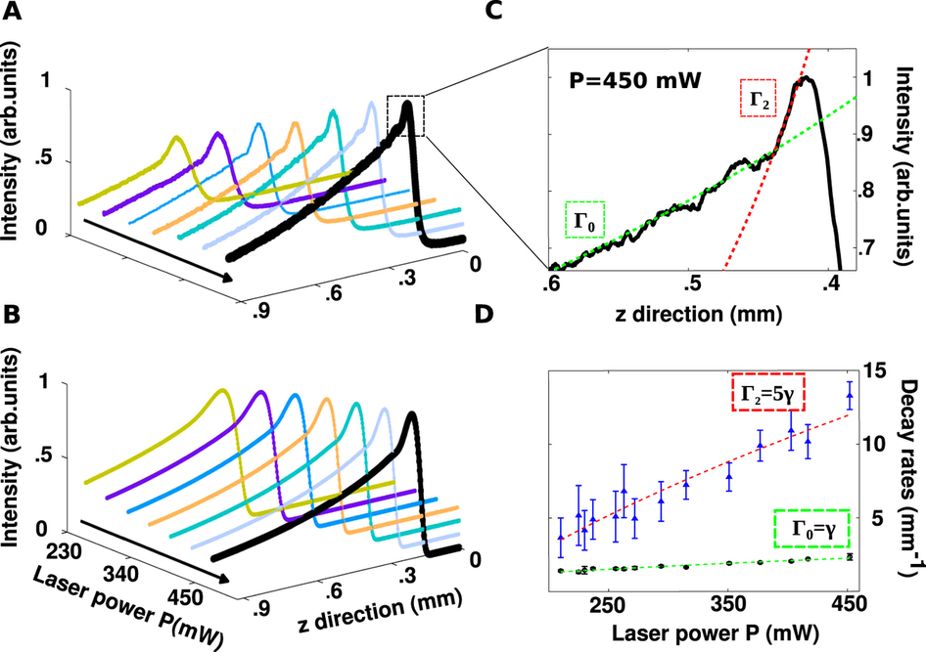}
\caption{
\textbf{A} Observed intensity decay at different laser powers, obtained by slicing along $X\simeq0.1$mm the top-view intensity distribution the propagation direction (see the yellow line in Fig.~\ref{fig7}B). \textbf{B} Numerically calculated decays in the conditions of panel A. \textbf{C} Peak region of the experimental curve at $P = 450$mW. The superposition of the first two exponential decays unveils the presence of two GVs, the fundamental state, $n=0$ (slowly decaying) and the first excited state, $n=2$ (fastly decaying). \textbf{D} Decay rates vs $P$ for the fundamental state, $\Gamma_0$ (filled circles) and the excited state, $\Gamma_2$, (triangles).

Reprinted by permission from Macmillan Publishers Ltd. from~\cite{2015GentiliniSciRep}. Copyright 2015.
\label{fig8}}
\end{figure}

Decay rates in Fig.~\ref{fig8} are detected by slicing the intensity profile $I(X,Z)$ at $X\simeq0.1$mm (yellow line in Fig.~\ref{fig7}B) and fitting the intensity versus $Z$ with two exponential functions. Different power levels exhibit very different dynamics.
The presence of double exponential decays, that is, the superposition of the first two GVs, is more evident at high power.
It was observed and calculated that double-exponential decay dynamics obey the quantized spectrum scaling $\Gamma_2/\Gamma_0=5$ at all investigated power levels, as shown in Fig.~\ref{fig8}D. 
This demonstrates that authors of~\cite{2015GentiliniSciRep} excited the fundamental state $\mathfrak{f}_0^-$ and the first excited state $\mathfrak{f}_2^-$. Odd states are not excited, as expected from Gaussian TEM$_{00}$ $x-$parity. Each of the two rates has a square root dependence on $P$, signature of the underlying nonlinearity. This power dependence distinguishes RHO dynamics from linear loss, due to absorption and scat\-te\-ring.

The RHO eigenstates are quasi-ei\-gen\-sta\-tes of the Fourier transform operator, which in optics represents the far field.
Let us consider the RHO Hamiltonian in the momentum basis ($\hat p \rightarrow \mathfrak{p}$ and $\hat x \rightarrow \imath\partial_{\mathfrak{p}}$)
\begin{equation}
\hat H_{RHO}(\mathfrak{p},i\partial_{\mathfrak{p}})=\frac{\mathfrak{p}^2}{2}+\frac{1}{2}\gamma^2\partial_\mathfrak{p}^2=-\hat H_{RHO}(-i\partial_x,x).
\end{equation}
Pure GVs are unfeasible to describe a physical experiment, because one cannot neglect that GVs have an infinite support, i.e., the $x$-region where the eigenfunction is not null, is not finite. In order to account for the spatial confinement of the experiment, authors of~\cite{2016Braidotti} introduced the windowed GVs:
\begin{equation}
\phi^W_G(x)=\sum_{n=0}^{N}\sqrt{\Gamma_n}\mathfrak{f}^-_n\langle\mathfrak{f}^+_n|\psi(x,0)\rangle\mbox{rect}_W(x),
\label{WGV}
\end{equation}  
where
$$\mbox{rect}_W(x)=\left\{\begin{array}{ccc}0&\mbox{for}|x|\geq W\\1&\mbox{for}|x|<W\end{array}\right.,$$
with $W$ is the finite size of the physical system.
During the evolution, the Gamow ground state has the lowest decay rate, i.e., $\gamma/2$. This allows to consider, in the long term evolution, only the fundamental GV, and so only the Fourier transform $\mathcal{F}$ of the fundamental state of Eq.~(\ref{WGV}):
\begin{equation}
\begin{array}{l}
\tilde \psi(k_x)=\mathcal{F}\left[\mathfrak{f}^{-,W}_0(x)\right]=\\
=\biggl(\frac{1}{4}+\frac{i}{4}\biggr)e^{-\frac{ikx^2}{2\gamma}}\frac{(-i\gamma \pi)^{1/4}}{W}\times\left\{-\mbox{Erf}\biggl[\frac{(\frac{1}{2}-\frac{i}{2})(k_x-W\gamma)}{\sqrt{\gamma}}\biggr]+\mbox{Erf}\biggl[\frac{(\frac{1}{2}-\frac{i}{2})(k_x+W\gamma)}{\sqrt{\gamma}}\biggr]\right\}.
\label{fftdn}
\end{array}
\end{equation}
Eq.~(\ref{fftdn}) provides an ana\-ly\-tical expression of the far field, which is compared below with the experiments. Indeed, Eq.~(\ref{fftdn}) allows to predict in closed form the typical M-shaped shock profile: it describes the internal undular bores and the correct sca\-ling of the undulation period with respect to the power, i.e. the period $T$ is predicted to scale with the square root of $\gamma$, and hence with the forth square root of the beam input power.

\begin{figure}[h!]
\centering
\includegraphics[width=0.95\textwidth]{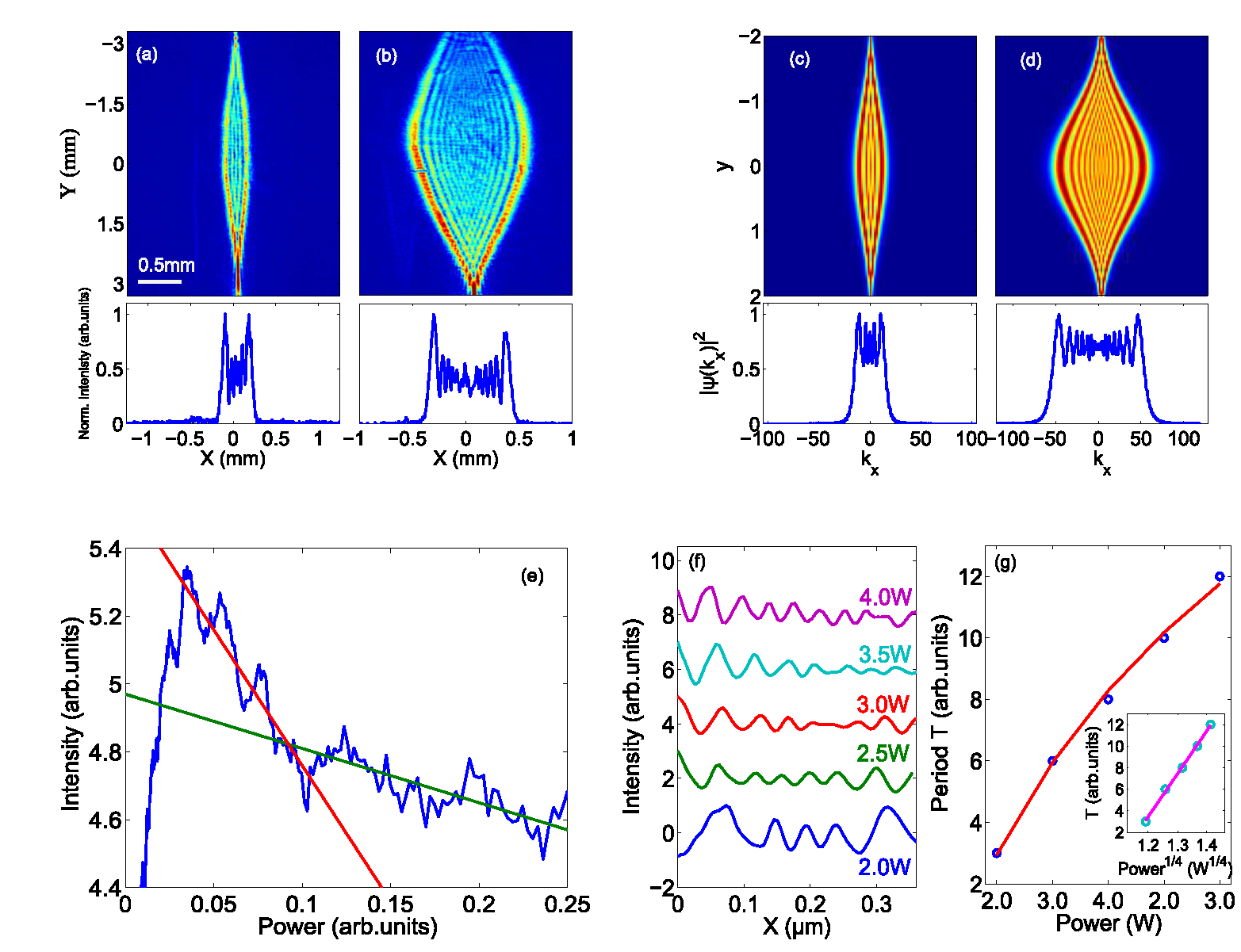}
\caption{
(a,b) CCD image of the light beam at laser powers $P = 2$W and $4$W, respectively; the bottom panels show the normalized intensity profile at the ma\-xi\-mum waist along $Y=0$. (c) Analytical solution obtained by Eq. (\ref{fftdn}) changing Gaussianly the power $P$ in the $y$ direction. (d) As in (c) but for higher powers; the bottom panels show the slice of panel (c) and (d) at $y = 0$, i.e., Eq.~(\ref{fftdn}) square modulus for $W = 1.5$ and  $\gamma \simeq 12$ and $\gamma\simeq 40$, respectively. (e)  Log-scale normalized intensity as a function of power, as obtained by slicing along $Y$ a region in panel (b). The slopes of the straight lines give the GV decay rates ($\gamma_1 = -8 \pm 0.4$ and $\gamma_2 = -1.6 \pm 0.1$). Their quantized ratio is $5.0 \pm 0.4$ as expected from theory~\cite{2015GentiliniPRA}. (f) Intensity oscillations for different power values. (g) Measured oscillations period $T$ as a function of power; continuous line is the fit function $T \propto \sqrt[4]{P}$, as expected by the theory; the inset shows the same curve of (g) with $P^{1/4}$ as abscissa axis.

Reprinted with permission from ~\cite{2016Braidotti}. Copyright 2016 Optical Society of America.
\label{fig9}}
\end{figure}

Figure~\ref{fig9} reports experimental results in RhB, through the previously described setup, and the comparison with the numerical results.
Images of the beam in the far field (corresponding to the square modulus of the spatial intensity Fourier transform) for different input powers were collected and shown in Figs.~\ref{fig9}a,b. For low power (not reported) the elliptical beam profile remains Gaussian along propagation. A different phenomenon occurs while increasing the power: the beam transverse section along $X$ broadens and develops intensity peaks on its lateral edges. Essentially, it becomes M-shaped.
These results are in remarkable agreement with Eq.~(\ref{fftdn}), as shown in Figs.~\ref{fig9}c,d.

Different positions in the $Y$ direction cor\-re\-spond to different power levels. Any power level fur\-ni\-shes a different value of $\gamma$, being $\gamma=\sqrt{\frac{p}{\sqrt{\pi}\sigma^2}}$.
The Gaussian beam profile in the $Y$ direction, that is, $p\propto exp(-y^2)$, provides the link between $Y$, $P$ and $\gamma$. 
This implies that, observing a CCD image, intensity profiles at dif\-fe\-rent $Y$ correspond to different powers.
Therefore, the expected exponential trend with respect to the power can be extracted from a single picture by looking at different $Y$ positions. 
Figure~\ref{fig9}e exhibits a fitting with two exponential decays in an intensity profile versus power. The extracted ratio of the related two decay rates is $5$ and hence in agreement with the expected quantized theoretical values described in Sec.~\ref{TAQM}.

Undular bores of DSWs were analyzed and exhibited in  Fig.~\ref{fig9}f, while the field intensity undulation period $T$ versus $P$ is shown in Fig.~\ref{fig9}g.
In order to demonstrate univocally that $T\propto\sqrt[4]{P}$, inset in Fig.~\ref{fig9}g reports the period $T$ as function of $\sqrt[4]{P}$. The resulting linear beha\-vior confirms the theoretical results.

\subsection{Nonlinearity and Disorder in Thermal Media}\label{disorder}

Thermal media have been investigated also in their interplay with disorder. Theoretical studies demonstrated that, even if solitons are stable under a certain amount of randomness, the latter competes with nonlinearity, while nonlocality filters disorder-induced scattering effects and soliton random walk can be efficiently suppressed in highly nonlocal media~\cite{2008Kartashov,2010Folli,2012Maucher}.
DSWs are nonlinear coherent oscillations, and the phenomenon of light scattering affects their formation in significant way~\cite{2012Ghofraniha}.

In this section we report experiments in two different optical systems that combines third-order nonlinearity (high-power laser beams) with nonlocality (thermal material response) and disorder (scattering particles). The first thermal medium is a dispersion of silica spheres of $1\mu$m diameter in $0.1$mM aqueous solution of RhB. The second one is a $1$mm$\times 1$mm$\times 8.5$mm parallelepiped of silica aerogel.
Despite observations of DSWs in disordered thermal media, a theoretical model that comprehends both nonlinearity, nonlocality and disorder has been developed only for solitons~\cite{2010Folli}. The existing theoretical model for DSWs is summarized below and neglects the nonlocality contribution. It approximates thermal nonlinearity to a local Kerr effect, and adds a random potential~\cite{2012Ghofraniha}.

We start from Eq.~(\ref{eq:NLS1}) with $\Delta n[|A|^2]=n_2|A|^2+\Delta n_R(X,Y,Z)$ and $L_{loss}\sim\infty$ (no loss). Through the same scaling of Sec.~\ref{kerr} and approximation to cylindrical symmetry $\partial_y\sim0$, we obtain
\begin{equation}
\imath\epsilon\partial_z\psi+\frac{\epsilon^2}2\partial_x^2\psi-|\psi|^2\psi+U_R\psi=0,
\label{eq:disorder}
\end{equation}
with $U_R(x,y,z)=\frac{\Delta n_R(X,Y,Z)}{n_2 I_0}$ taken as a random dielectric noise mainly acting on the phase~\cite {2012Ghofraniha}.
In the hydrodynamic limit $\epsilon\sim0$, the phase chirp behaves like a moving unitary mass particle~\cite{2012Ghofraniha}:
\begin{equation}
\frac{\mbox{d}^2x}{\mbox{d}z^2}=-\frac{\mbox{d}U}{\mbox{d}x}+\eta_R,
\label{eq:randomchirp}
\end{equation}
with $U=\exp\left(-x^2/2\right)$ the deterministic potential for a Gaussiam TEM$_{00}$ given by the nonlinearity, and $\eta_R=-\frac{\mbox{d}U_R}{\mbox{d}x}$ a Langevin force with Gaussian distribution, such that $\langle\eta_R(z)\eta_R(z')\rangle=\eta^2\delta(z-z')$ and $\eta=\sqrt{\left\langle\left(\frac{\mbox{d}U_R}{\mbox{d}x}\right)^2\right\rangle}\simeq\sqrt{\langle\left(\Delta n_R\right)^2\rangle}(|n_2|I_0)^{-1}$ the disorder strength. Brackets $\langle,\rangle$ denote the statistical average, and the dependence of $\eta_R$ on $x,y$ is neglected for stochastic independence and cylindrical symmetry, respectively, thus $\eta_R\simeq\eta_R(z)$.

\begin{figure}[h!]
\centering
\includegraphics[width=0.65\textwidth]{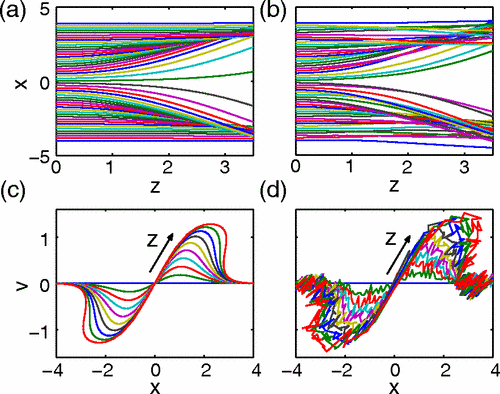}
\caption{
(a,b) Trajectories $x(z)$ and (b,d) phase space $(x,v)$, respectively with disorder strength $\eta=0$ and $\eta=0.1$. $z$ varies from $z=0$ to $z=3$.

FIG.~4 reprinted with permission from~\cite{2012Ghofraniha}. Copyright 2012 by the American Physical Society.
\label{fig11}}
\end{figure}

Figure~\ref{fig11} shows trajectories $x(z)$ (Figs.~\ref{fig11}a,b) and phase space $(x,v)$ (Figs.~\ref{fig11}c,d), where $v=\frac{\mbox{d}x}{\mbox{d}z}$, respectively without ($\eta=0$) and with ($\eta=0.1$) disorder, the latter obtained by a stochastic Runge-Kutta algorithm~\cite{1992Honeycutt1,1992Honeycutt2}.
In absence of disorder (Figs.~\ref{fig11}a,c) the shock is signaled by the intersection of multiple trajectories $x(z)$ and, in the phase space, this corresponds to the induced wave breaking phenomenon, that is, the folding of the velocity profile into a multivalued function for increasing $z$.
In presence of disorder, Figs.~\ref{fig11}b,d, the particle-like dynamics tends to diffuse, as is evident from the related trajectories and phase space. Correspondingly, the propagation distance before the intersections is greater for the disordered case and the shock is delayed in the $z$ direction.

\begin{figure}[h!]
\centering
\includegraphics[width=0.7\textwidth]{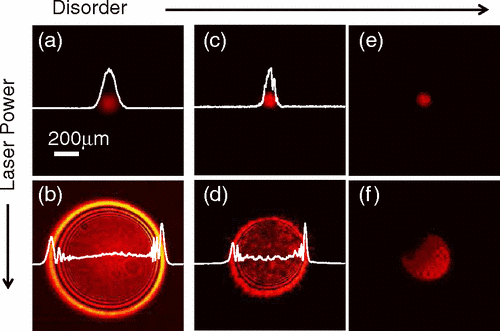}
\caption{
Transverse intensity patterns for different input power $P$ and silica spheres concentration $c$: (a) $P=5mW$, $c=0$w$/$w, (b)$P=400mW$, $c=0$w$/$w, (c) $P=5mW$, $c=0.017$w$/$w, (d) $P=400mW$, $c=0.017$w$/$w, (e) $P=5mW$, $c=0.030$w$/$w, (f) $P=400mW$, $c=0.030$w$/$w. White 1D curves show the measured section of the intensity profiles vs $X$.

FIG.~1 reprinted with permission from~\cite{2012Ghofraniha}. Copyright 2012 by the American Physical Society.
\label{fig10}}
\end{figure}

We report here the experiments in RhB with silica spheres dispersions~\cite{2012Ghofraniha}.
In order to vary the degree of disorder, several silica concentrations were prepared, ranging from $0.005$w$/$w to $0.03$w$/$w, in units of weight of silica particles over suspension weight.
The experimental setup is similar to that illustrated in Fig.~\ref{fig7}A. The first lens focuses the beam on the input facet of the sample, reaching a beam waist $W_0\simeq10\mu$m. The aqueous solutions are put in $1$mm$\times1$cm$\times3$cm glass cells with propagation along the $1$mm vertical direction (parallel to gravity) to moderate the effect of heat convection.
All measurements are performed after the temperature gradient has reached the stationary state and the particle suspensions are completely homogeneous.
In~\cite{2012Ghofraniha}, main loss mechanisms are absorption and scattering. The measured loss length (absorption plus scattering) varies in a range from $1.2$mm to $1.6$mm (highest value is for for pure dye solution). These values are obtained by fitting with exponential decay the beam intensity vs propagation distance $Z$. The fact that the loss length is always greater than the position of the shock point~\cite{2012Ghofraniha} allowed authors to neglect losses at a first approximation in their theory. In addition, they found that the scattering mean free path is of the order of millimeters for all the considered samples. In Fig.~\ref{fig10} images of the transmitted beam on the transverse plane for different input laser powers $P$ and various concentrations $c$ are shown. The number and the visibility of the DSW oscillations increase with $P$ and decrease with $c$, evidence of DSWs enhancement by nonlinearity and inhibition by disorder.

\begin{figure}[h!]
\centering
\includegraphics[width=0.85\textwidth]{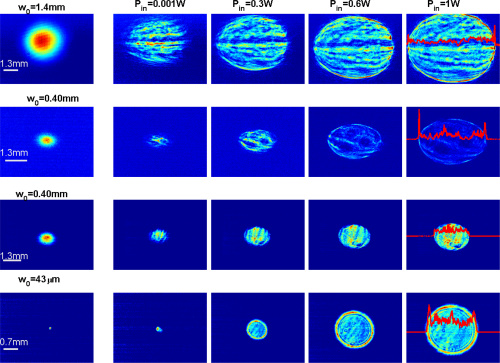}
\caption{
 Far field intensity profiles at the output of the silica aerogel for $P_{in}$ ranging from $1$mW to $1$W, and input beam waist $w_0$ ranging from $43\mu$m to $1.4$mm. Images in the second and third rows correspond to the same incident laser power and beam size, but different positions of the incident laser beam.

Reprinted with permission from~\cite{2014Gentilini}. Copyright 2014 Optical Society of America.
\label{fig12}}
\end{figure}

Experimental observations have been also performed in silica aerogel~\cite{2014Gentilini}.
The silica aerogel samples are prepared following a base-catalyzed sol-gel procedure~\cite{1998Venkateswara}, and in-depth details are given in~\cite{2014Gentilini}.
It turns out that the sample used in the experiment has mass density $\rho=0.215$g$/$cm$^3$ and refractive index $n_0=1.074$. 
Experimental setup is very similar to the previous ones (Fig.~\ref{fig1}a), except for the sample.
In~\cite{2014Gentilini}, authors vary the input beam waist $W_0$, the input laser power $P_{in}$, and record the transmitted intensity distribution $I(X,Y,Z=8.5\mbox{mm})$ by the CCD camera. Observations are shown in Fig.~\ref{fig12}. Images in the second and third rows of Fig.~\ref{fig12} correspond to the same experimental conditions in term of incident laser power and beam size, but the incident laser beam impinges on different points. In correspondence of regions of the silica aerogel sample displaying low enough disorder (second row), a transition from scattering dominated regimes to nonlinear regimes is present: at moderate powers DSWs are not observed because of scattering losses, at high powers DSWs can be generated.

\subsection{Dispersive Shock Waves in Biological Suspensions and Chemical Compounds}\label{bio}

The study of optical effects in light propagation through chemical and biological solutions is a field of growing interest~\cite{1982Latimer,1987Ashkin,2014Smith,2014Smith1,2017Bezryadina,2017Dervaux,2019Gautam}, both from a linear and a nonlinear perspective.
However, although observations of nonlinear optical phenomena in chemical and soft-matter systems can be found in a large literature~\cite{2004Conti,2005Conti,2007Ghofraniha,2007El-Ganainy,2012Ghofraniha,2013Gentilini,2013Greenfield,2013Man,2014Gentilini,2014Smith,2015GentiliniSciRep,2016Braidotti}, and new experiments in chemical media are useful only if the material owns very specific properties, little is known about nonlinearity in biological fluids and the related literature is very recent~\cite{2017Bezryadina, 2019Gautam}.
Bio-materials can be very interesting, because both chemical and biological compounds can be excellent tunable thermal media, and DSWs were already observed~\cite{2007Ghofraniha,2014Smith1,2019Gautam}. 

For sake of completeness, in this section we report two experiments. The first one is in M-Cresol/Nylon, a chemical solution that exhibits an isotropic giant self-defocusing nonlocal nonlinearity, tunable by varying the nylon concentration~\cite{2014Smith1}. The second one is in human red blood cell suspensions, where the concentration of hemoglobin~(Hb) and the input laser beam power make the nonlinearity change from self-focusing to nonlocal defocusing~\cite{2019Gautam}.

\begin{figure}[h!]
\centering
\includegraphics[width=0.85\textwidth]{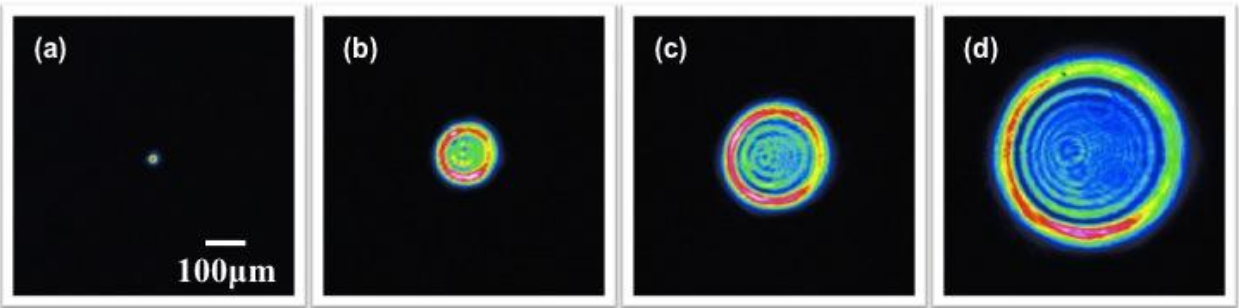}
\caption{
Output beam intensity transverse profiles, coming out from a $2$mm long M-Cresol/Nylon solution. Input power varies: (a) $P_{in}=2\mu$W, (b) $P_{in}=5$mW, (c) $P_{in}=10$mW, (d) $P_{in}=20$mW.

Reprinted with permission from~\cite{2014Smith1}. Copyright 2014 Optical Society of America.
\label{fig14}}
\end{figure}

Figure~\ref{fig14} shows transverse profiles of output beam intensity after a propagation of $2$mm in M-Cresol/Nylon.
M-Cresol/Nylon is made up of an organic solvent (m-cresol) and a synthetic polymeric solute (nylon). When it is enlightened by a CW laser beam, light absorption induces local temperature variations, which reduces the refractive index, that is, the material experiences a nonlinear thermo-optical effect. In particular, \cite{2014Smith}'s authors measured  the M-Cresol/Nylon nonlinear Kerr coefficient $n_2$ and found that, if for pure m-cresol it is $-9\times10^{-8}$cm$^2/$W, for a nylon mass concentration of $3.5\%$ it is $-1.6\times10^{-5}$cm$^2/$W, an order of magnitude higher than most of the other thermal nonlinear materials reported in literature. Authors generated the DSWs in Fig.~\ref{fig14} by focusing the input beam (a CW laser of wavelength $532$nm) to $20\pm1\mu$m onto the surface of M-Cresol/Nylon solution of $3.5\%$ nylon concentration. The input laser power was  varied ranging from $2\mu$W to $20$mW and, when it reached $5$mW, the  wave-breaking occured.

\begin{figure}[h!]
\centering
\includegraphics[width=0.75\textwidth]{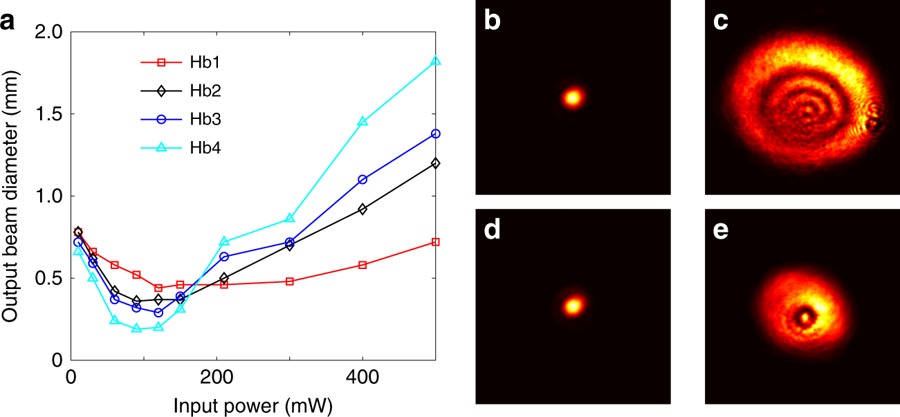}
\caption{
Output beam waist for varying hemoglobin concentration and input power.
\textbf{a} Detected beam diameter as function of input power through the hemoglobin solutions for four different concentrations (Hb1-Hb4): $2.4$, $5.1$, $8.6$, and $15.0$ million cells per mL.
Nonlinear self-focusing of the beam occurs around $100$mW for high concentrations of hemoglobin, but it subsequently expands into thermal defocusing rings at high powers.
\textbf{b-e} Output beam transverse intensity profiles for \textbf{b} self-trapped beam at high concentration and low power, \textbf{c} DSW at high concentration and high power, \textbf{d} self-trapped beam at low concentration and low power, \textbf{e} DSW at low concentration and high power.

Reprinted by permission from Macmillan Publishers Ltd. from~\cite{2019Gautam}. Copyright 2019.
\label{fig13}}
\end{figure}

Figure~\ref{fig13} reports a part of the results obtained in lysed human red blood cells aged samples, where free Hb determines sign and nonlocality of the optical nonlinearity from self-focusing (and self-trapping) to strong thermal defocusing effects, regime in which DSWs occur~\cite{2019Gautam}. 
Beyond the biological issues related to human red blood cells, holding uncountable applications to life sciences and medicine, red blood cells refractive index tunability makes this medium be incredibly interesting also from a physical point of view~\cite{1988Steinke,2006Ghosh,2015Miccio,2018Gautam}. In normal conditions, red blood cells are disc-shaped malleable cells, averagely with $8\mu$m of diameter and $2\mu$m of thickness, which have a spatially uniform refractive index because of the lack of nuclei and most organelles~\cite{2006Ghosh,2018Gautam}. To enable the passage through veins and narrow  microcapillaries, red blood cells  exhibit  distinctive deformability. Since their optical properties depend on the shape and refractive index of cells, they can be used as tunable optofluidic microlenses~\cite{2015Miccio}.

The red blood cell refractive index is mainly determined by Hb, which is the largest part of the erythrocyte dry content by weight~\cite{1988Steinke}. Fig.~\ref{fig13}a shows the output beam waist as a function of input power through the Hb solutions for four different concentrations, from $2.4$ to $15.0$ million cells per mL.
Experiments in~\cite{2019Gautam} are performed by using a linearly polarized CW laser beam with a wavelength of $532$nm focused through a lens of $125$mm focal length into a $3$cm long glass cuvette filled with the red blood cell suspensions. In particular, the focused beam has initial waist $W_0=28\mu$m at the focal point, which was located at $1$cm away from the input facet of the cuvette to avoid heating and surface effects~\cite{2017Bezryadina}. Outputs from the sample were monitored with a CCD camera and a power detector, and are reported in Figs.~\ref{fig13}b-e, at variance of Hb concentration and input power. DSWs occur at high power (Figs.~\ref{fig13}c,e), more visible in high Hb concentration regime (Fig.~\ref{fig13}c).

\subsection{Incoherent Dispersive Shock Waves}\label{incoherent}

In this section, we report experimental evidence of incoherent DSWs theoretically introduced in Sec.~\ref{turbulence}. In particular, we show observations of the transition from shocklets to collective incoherent DSWs~\cite{2015Xu}.

\begin{figure}[h!]
\centering
\includegraphics[width=0.75\textwidth]{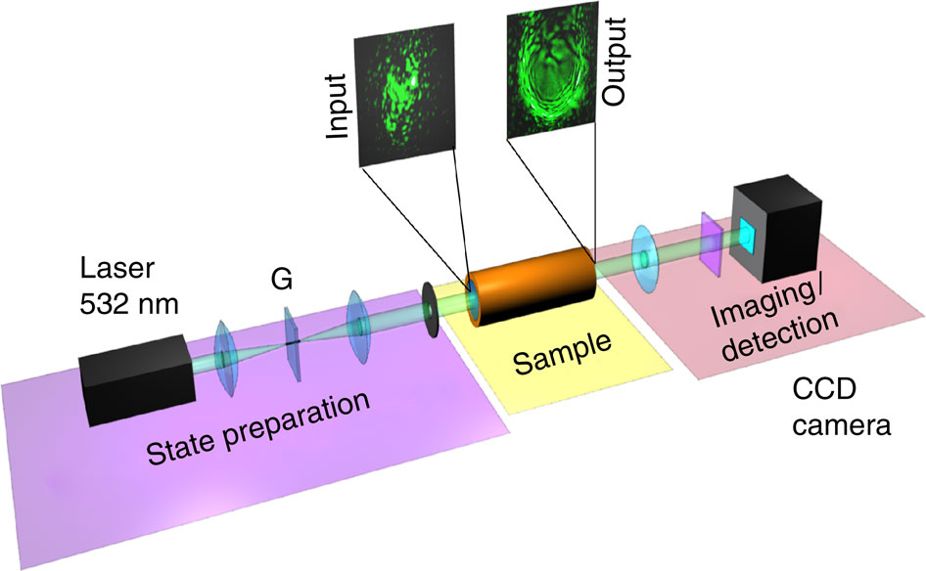}
\caption{
Experimental setup. A CW laser beam, $\lambda=532$nm, is sent through a 4-f telescope. A ground-glass plate (G), placed in the midst of the telescope (on the focus of the first lens), generates a speckle pattern. The incoherent beam impinges the samples, a cylindrical tube filled with a solution of methanol and graphene nanoscale flakes, with waist $W_0=2.3$mm, while the initial coherence length $\lambda_c^0$ is controlled by changing the beam size on G. A CDD camera detects the output.

Reprinted by permission from Macmillan Publishers Ltd. from~\cite{2015Xu}. Copyright 2015.
\label{fig15}}
\end{figure}

By varying the effective range of nonlocality, authors of~\cite{2015Xu} performed experiments both in the quasi-local and in the highly nonlocal regime. The experimental setup is sketched in Fig.~\ref{fig15}. A CW laser ($\lambda=532$nm) is made incoherent by passing through a 4-f telescope with a ground-glass plate in the middle. The initial coherence length $\lambda_c^0\sim 200\mu$m, that is, the size of the speckles induced in the beam at the sample input facet, is controlled by changing the waist impinging on the ground-glass plate.
The sample is a dilute solution of methanol and graphene nanoscale flakes. The latter provide optimal conversion of absorbed laser energy into heat, thanks to the absence of fluorescence mechanisms and a negligible absorption. A CCD camera detects intensity images at the output.

\begin{figure}[h!]
\centering
\includegraphics[width=0.95\textwidth]{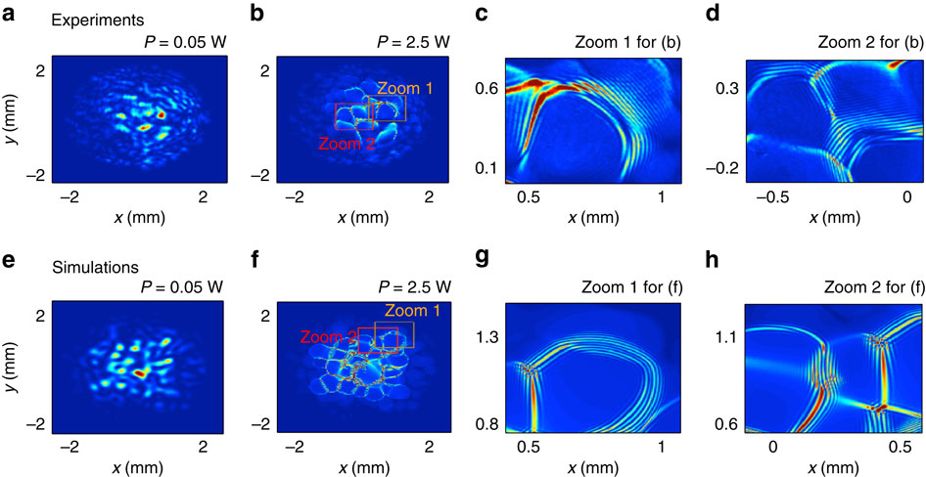}
\caption{
Experimental observation at short-range regime. \textbf{a,b} Experimental beam profiles of the output intensity recorded at power \textbf{a} $P=0.05$W, \textbf{b} $P=2.50$W. \textbf{c,d} Zooms on details of (b) that evidence the development of shocklets. \textbf{e,f} Numerical simulations of NLSE equation, and \textbf{g,h} corresponding zooms.

Reprinted by permission from Macmillan Publishers Ltd. from~\cite{2015Xu}. Copyright 2015.
\label{fig17}}
\end{figure}

Figure~\ref{fig17} shows the shocklets formation in a quasi-local regime. To reach this stage, concentration of graphene nano-flakes was increased, to increase the absorption and reduce the nonlocality, and a higher coherence length was chosen ($\lambda_c^0\sim 250\mu$m), to inhibit collective behaviors. At low power (Fig.~\ref{fig17}a for experiments, Fig.~\ref{fig17}e for NLSE simulations) propagation is linear, so dominated by diffraction. As the input power increases (Fig.~\ref{fig17}b for experiments, Fig.~\ref{fig17}f for NLSE simulations), each speckle develops its own DSW, with a regular undular pattern. These is more evident in experimental (Figs.~\ref{fig17}c,d) and numerical (Figs.~\ref{fig17}g,h) zooms of shocklets undular bores.

\begin{figure}[h!]
\centering
\includegraphics[width=0.95\textwidth]{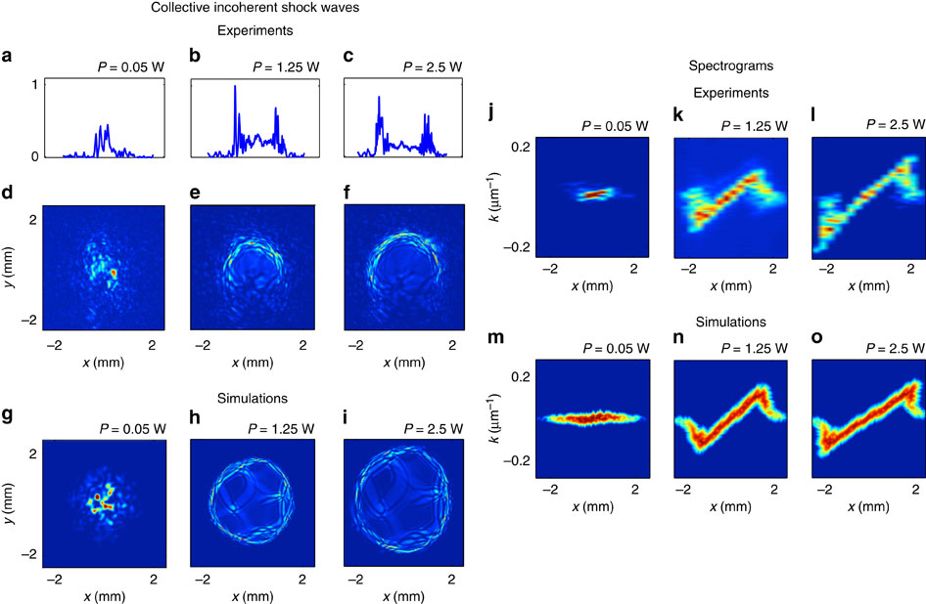}
\caption{
Experimental observation at long-range regime. \textbf{a-c} One dimensional intensity transverse profiles along $x$ at $y=0$. \textbf{d-f} Two dimensional intensity profiles. The asymmetry in the lower part of the beam is due to convection within the sample. \textbf{g-i} Numerical simulations of NLSE. \textbf{j-l} Experimental and \textbf{m-o} numerical spectrograms: the Z-shaped distortion reveals a dramatic coherence degradation on the annular boundaries of the beam (the coherence length decreases at the shock front), while a significant coherence enhancement occurs in the internal region of the beam. Input beam power: (a,d,g,j,m) $P=0.05$W, (b,e,h,k,n) $P=1.25$W, (c,f,i,l,o) $P=2.50$W.

Reprinted by permission from Macmillan Publishers Ltd. from~\cite{2015Xu}. Copyright 2015.
\label{fig16}}
\end{figure}

Figure~\ref{fig16} reports the emergence of a giant collective incoherent DSW in highly nonlocal regime. Such a transition was made by decreasing the concentration of graphene nano-flakes and the coherence length ($\lambda_c^0\sim 200\mu$m). At low power no collective behavior was observed, neither in the intensity profile (Figs.~\ref{fig16}a,d for experiments, Fig.~\ref{fig16}g for NLSE simulations), nor in the spectrogram (Fig.~\ref{fig16}j for experiments, Fig.~\ref{fig16}m for LRVE simulations).
At nonlinear regime, the occurrence of the annular reshaping of the beam is visible, with high frequencies piled on the boundaries, and low frequencies dominating the central part (Figs.~\ref{fig16}b,c,e,f for experiments, Figs.~\ref{fig16}h,i for NLSE simulations). The corresponding spectrograms show the linear behavior (Figs.~\ref{fig16}j,m) at low power and Z-shape distortion at higher powers. The latter phenomenon is a signature of incoherent DSWs (see Sec.~\ref{turbulence}), and is observable in Figs.~\ref{fig16}k,l and Figs.~\ref{fig16}n,o, respectevely in experiments and simulations.

All the shock phenomena reported here, as DSWs in general, do not arise in the weakly nonlinear (weak turbolence) regime, but solely in the strongly nonlinear (strong turbulence) regime. The Vlasov equation is valid beyond the weakly nonlinear regime and thus describes the collective incoherent shocks in the highly nonlocal case~\cite{2015Xu}. However, in the weakly nonlocal regime, there is no theory that describes the development of the shocklets reported in Fig.~\ref{fig17}: despite intense efforts from several decades, strong turbulence still constitutes a challenging unsolved problem of classical physics.

\section{Conclusions}\label{concl}
We reviewed the most widespread current theoretical models that describe nonlocal NLSE DSWs in spatial optical beam propagation.
Moreover, we discussed their experimental observations.

In Sec.~\ref{nonlocalnlse} the derivation of nonlocal NLSE was detailed, and main features of wave breaking in thermal Kerr media were reported~\cite{2007Ghofraniha}.
In order to exhibit the theoretical interpretations of these phenomena as intrinsically irreversible, TAQM and turbulence wave theory approaches were summarized~\cite{2017Marcucci,2015Xu,2016Xu}.

Section~\ref{exp} is a collection of experiments on DSW generation in thermal media, first about a quite rich literature on observations in Rhodamine~\cite{2007Ghofraniha}, and their TAQM explaination~\cite{2015GentiliniPRA,2015GentiliniSciRep,2016Braidotti}. As second instance, we analyzed the interaction between disorder and nonlinearity in Rhodamine with silica spheres~\cite{2012Ghofraniha} and in silica aerogel~\cite{2014Gentilini}, where the randomness inhibits the DSWs occurrence. Moreover, we reviewed very recent works on generation of photonic wave breaking in chemical~\cite{2014Smith1} and biological solutions~\cite{2019Gautam}, fields where DSWs are emerging as surprising tools, useful for sensing and control of extreme phenomena. Finally, we showed emergence of giant collective incoherent shock waves from random wave propagation in highly nonlocal media~\cite{2015Xu}.

The study of nonlinear optics in new materials, like soft matter and biological suspensions, is opening the way to a new branch of photonics. Many applications include, but are not limited to, spectroscopy, medicine, life sciences, non invasive diagnosis and time-resolved low-power probes. May the physics of nonlinear waves support the development of these new directions? Can novel mathematical tools deepen our understanding of nonlinear radiation-matter interaction?
This manuscript is intended to sustain the improvement of theory and experiments concerning nonlinear optical propagation in highly nonlocal and complex matter.

We thank A. Picozzi, S. Trillo and G. Xu for the thoughful discussions and suggestions on the manuscript, and MD Deen Islam for the assistance in the laboratory. We acknowledge support from the QuantERA ERA-NET Co-fund 731473 (Project QUOMPLEX), H2020 project grant number 820392, Sapienza Ateneo, PRIN 2015 NEMO, PRIN 2017 PELM, Joint Bilateral Scientic Cooperation CNR-Italy/RFBR-Russia 2018-2020, and the National Key R\&D Program of China (2017YFA0303800).

\bibliographystyle{unsrt}
\bibliography{references}

\end{document}